\documentclass[11pt]{article}
\usepackage{epsfig}
\usepackage{latexsym}
\usepackage{graphicx}
\usepackage{color}
\usepackage{amsmath,amssymb}
\usepackage{cite}

\makeatletter
\def\section{\@startsection {section}{1}{\z@}{-3.5ex plus -1ex minus
 -.2ex}{2.3ex plus .2ex}{\large\bf}}
\def\subsection{\@startsection{subsection}{2}{\z@}{-3.25ex plus -1ex
minus -.2ex}{1.5ex plus .2ex}{\normalsize\bf}}
\makeatother
\newcommand{\captionfonts}{\small}
\makeatletter  % Allow the use of @ in command names
\long\def\@makecaption#1#2{%
  \vskip\abovecaptionskip
  \sbox\@tempboxa{{\captionfonts #1: #2}}%
  \ifdim \wd\@tempboxa >\hsize
    {\captionfonts #1: #2\par}
  \else
    \hbox to\hsize{\hfil\box\@tempboxa\hfil}%
  \fi
  \vskip\belowcaptionskip}
\makeatother   % Cancel the effect of \makeatletter
\catcode`@=11
\def\marginnote#1{}
\newcount\hour
\newcount\minute
\newtoks\amorpm
\hour=\time\divide\hour
by60
\minute=\time{\multiply\hour by60 \global\advance\minute
by-\hour}
\edef\standardtime{{\ifnum\hour<12 \global\amorpm={am}
\else\global\amorpm={pm}\advance\hour by-12 \fi
 \ifnum\hour=0
\hour=12 \fi
 \number\hour:\ifnum\minute<10
0\fi\number\minute\the\amorpm}}
\edef\militarytime{\number\hour:\ifnum\minute<10
0\fi\number\minute}
\def\draftlabel#1{{\@bsphack\if@filesw
{\let\thepage\relax
 \xdef\@gtempa{\write\@auxout{\string
\newlabel{#1}{{\@currentlabel}{\thepage}}}}}\@gtempa
 \if@nobreak
\ifvmode\nobreak\fi\fi\fi\@esphack}
\gdef\@eqnlabel{#1}}
\def\@eqnlabel{}
\def\@vacuum{}
\def\draftmarginnote#1{\marginpar{\raggedright\scriptsize\tt#1}}
\def\draft{\oddsidemargin
0.0truein
 \def\@oddfoot{\sl preliminary draft \hfil
\rm\thepage\hfil\sl\today\quad\militarytime}
 \let\@evenfoot\@oddfoot
\overfullrule 3pt
 \let\label=\draftlabel
\let\marginnote=\draftmarginnote
\def\@eqnnum{(\theequation)\rlap{\kern\marginparsep\tt\@eqnlabel}
\global\let\@eqnlabel\@vacuum}
}
\catcode`@=12
\def\dj{\hbox{d\kern-0.347em \vrule width 0.3em height 1.252ex depth
-1.21ex \kern 0.051em}}

\def\ee{{\rm e}\,}

\def\Dirac{\,\raise.15ex\hbox{/}\mkern-13.5mu D}
\def\dirac{\,\raise.15ex\hbox{/}\kern-.57em \partial}
\def\aslash{\,\raise.15ex\hbox{/}\mkern-13.5mu A}
\def\shalf{{\ifinner {\textstyle \frac{1}{2}}\else \frac{1}{2} \fi}} 
\def\sthreehalfs{{\ifinner {\textstyle \frac{3}{2}}\else \frac{3}{2} \fi}} 
\def\sshalf{{\ifinner {\scriptstyle \frac{1}{2}}\else \frac{1}{2} \fi}} 
\def\sfourth{{\ifinner {\textstyle \frac{1}{4}}\else frac{1}{4} \fi}}
\def\sphifour{{\ifinner {\textstyle \frac{1}{4!}}\else \frac{1}{4!} \fi}}
\def\lsim{\stackrel{<}{_\sim}}

\def\XXint#1#2#3{{\setbox0=\hbox{$#1{#2#3}{\int}$}
     \vcenter{\hbox{$#2#3$}}\kern-.5\wd0}}

\def\bea{\begin{eqnarray}} \def\eea{\end{eqnarray}}
\def\be{\begin{eqnarray}} \def\ee{\end{eqnarray}} 
\newcommand{\promille}{%
  \relax\ifmmode\promillezeichen
        \else\leavevmode\(\mathsurround=0pt\promillezeichen\)\fi}
\newcommand{\promillezeichen}{%
  \kern-.05em%
  \raise.5ex\hbox{\the\scriptfont0 0}%
  \kern-.15em/\kern-.15em%
  \lower.25ex\hbox{\the\scriptfont0 00}}
\newcommand{\beq}{\begin{eqnarray}}
\newcommand{\eeq}{\end{eqnarray}}

\newcommand{\gsim}{\raisebox{-0.13cm}{~\shortstack{$>$ \\[-0.07cm]
      $\sim$}}~}
\newcommand{\s}{\newline \vspace*{-3.5mm}}
\begin{document}

\thispagestyle{empty}

\begin{center}
\hfill KA-TP-37-2010

\begin{center}

\vspace{1.7cm}

{\LARGE\bf Composite Higgs Boson Pair Production at the LHC}
\end{center}

\vspace{1.4cm}

{\bf R. Gr\"ober$^{\,a}$} and {\bf M. M\"uhlleitner$^{\,a}$}\\

\vspace{1.2cm}

${}^a\!\!$
{\em {Institut f\"ur Theoretische Physik, Karlsruhe Institute of Technology, 76128 Karlsruhe, Germany}
}\\

\end{center}

\vfill

\centerline{\bf Abstract}
\vspace{2 mm}
\begin{quote}
\small
The measurement of the trilinear and quartic Higgs self-couplings
is necessary for the reconstruction of the Higgs potential. This way
the Higgs mechanism as the origin of electroweak symmetry breaking can
be tested. The couplings are accessible in multi-Higgs production
processes at the LHC. In this paper we investigate the prospects of
measuring the trilinear Higgs coupling in composite Higgs models. 
In these models, the Higgs boson emerges as a pseudo-Goldstone boson
of a strongly interacting sector, and the Higgs potential is generated
by loops of the Standard Model (SM) gauge bosons and fermions. The
Higgs self-couplings are
modified compared to the SM and controlled by the compositeness
parameter $\xi$ in addition to the Higgs boson mass. We construct
areas of sensitivity to the trilinear Higgs coupling in the relevant
parameter space for various final states. 
\end{quote}

\vfill

\newpage
%
%%%%%%%%%%%%%%%%%%%%%%%%%%%%%%%%%%%%%%%%%%%%%%%%%%%%%%%
\section{Introduction}
\label{sec:Intro}
The Standard Model of particle physics has been extremely
successful in describing the fundamental forces of the weak,
electromagnetic and strong interactions. 
It is based on the $SU(2)_L\times U(1)_Y$ gauge group
which is spontaneously broken down to the electromagnetic group
$U(1)_{em}$. The electroweak symmetry breaking (EWSB) mechanism has been
implemented in the most economical way by adding a single weak 
isodoublet scalar field \cite{Higgs,Goldstone,Gunion}. 
Three of its four degrees of freedom, the
pseudo-Nambu-Goldstone bosons, provide the masses of the massive
electroweak gauge bosons, $W^\pm$ and $Z$. The remaining physical
degree of freedom is associated with the SM Higgs boson. Also the
fermions acquire their masses 
through the interaction with the Higgs boson in the ground state. The
non-vanishing vacuum expectation value, which is essential for the non-zero
particle masses, is induced by the typical form of the Higgs
potential. Since all the Higgs couplings are predetermined, the
parameters describing the Higgs particle are entirely fixed by its
mass. It is the only unknown parameter in the SM Higgs sector
\cite{Gunion}. Furthermore, the Higgs boson plays the role of an
ultraviolet moderator. It ensures unitarity in the scattering
of massive longitudinal gauge bosons. \s

Despite the fact that no experimental evidence has been found so far
for the Higgs mechanism, the SM is in very good agreement with
electroweak precision tests at LEP, SLC, Tevatron and HERA. Any
departure from the SM has to pass the test of EW precision
measurements. Considering new physics beyond the minimal version
of the SM Higgs hence motivates smooth departures from the SM. This is
{\it i.e.} given by the introduction of a light Higgs
boson which emerges as a pseudo-Goldstone boson from a
strongly-coupled sector \cite{Contino:2010rs}, the so-called Strongly
Interacting Light Higgs 
(SILH) scenario~\cite{Giudice:2007fh,Contino:2010mh}. It is a bound
state emerging from strong dynamics~\cite{Kaplan:1983fs,
  othercompositeHiggs}, and 
due to its Goldstone nature it is separated by a mass gap from the 
other usual resonances of the strong sector. 
At low energy, the composite model therefore has the same
particle content as in the SM, with a light and narrow Higgs-like scalar.
Because of the composite nature its couplings, however, are different from
the SM case. This can have significant impacts on the experimental
sensitivities in the LHC Higgs boson search channels
\cite{Giudice:2007fh,Falkowski:2007hz,Espinosa:2010vn}. \s

In Ref.~\cite{Giudice:2007fh} an effective Lagrangian for a composite
light Higgs boson was constructed. It was shown that with respect to
LHC studies the Higgs properties are described by its mass and two new
parameters. The effective SILH Lagrangian represents the first term of
the expansion in
$\xi = (v/f)^2$ where $v=1/\sqrt{\sqrt{2}G_F} \approx 246$ GeV is the
scale of EWSB and $f$ is the strong dynamics scale. It describes the
physics near the SM limit $\xi \to 0$, whereas in the technicolor
limit \cite{Hill:2002ap}, $\xi \to 1$, a resummation of the full series in $\xi$ is
needed. Such a resummation is provided for instance by explicit models built in
five-dimensional warped space. In this paper we chose
two five-dimensional models which we hope to be representative of
minimal composite Higgs models.
The deviations from the SM couplings are governed by only one
parameter $\xi $ which varies from 0 to 1. 
We did not take into account couplings with a different Lorentz structure. 
They are generated via the exchange of heavy resonances of
the strong sector and suppressed by at least $(f/m_\rho)^2$, with
$m_\rho> 2.5$~TeV being the typical scale of the heavy
resonances. A direct Higgs coupling to two gluons or two photons will 
also always be sub-leading compared to the couplings considered in this
paper. And the effect of new heavy top partners is already included in
the effective Lagrangian approach
\cite{Giudice:2007fh,Falkowski:2007hz,Espinosa:2010vn,Low:2010mr}. \s 

While the SM Higgs boson suffers from the hierarchy problem, this
problem does not arise for a composite Higgs state. The Higgs
potential vanishes at tree level due to the non-linear Goldstone
symmetry acting on it. However, the global symmetry of the strong
sector is explicitly broken by the couplings of the SM fields to the
strong sector. The Higgs potential is thus generated by loops of SM
fermions and gauge bosons. The EWSB scale is dynamically
generated and can be smaller than the scale $f$, in contrast to
technicolor models where there is no separation of scales. The Higgs
gets a light mass through loops with $m_h \sim g_{SM} v$ where
$g_{SM} \lsim 1$ is a generic SM coupling. \s

Although the composite Higgs boson couplings, deviating from the
SM, are no direct probe of the strong sector at the origin of the
electroweak symmetry breaking\footnote{A direct probe is provided by
  the production of heavy resonances of the strong sector. Further
  tests are given by the observation of longitudinal gauge boson
  scattering growing with the energy or strong double Higgs production
  in longitudinal gauge boson fusion \cite{Contino:2010mh}. See also
  Ref. \cite{Accomando:2007xc} for a probe of EWSB.} their determination 
would allow for first insights in the dynamics
controlling the Higgs sector. With the measurement of the Higgs
self-couplings the Higgs potential can the reconstructed and
thus the Higgs sector of EWSB can be tested. Furthermore, consistency
tests within the framework of the considered model can be performed by
comparing with the other Higgs couplings to fermions and gauge bosons. 
In this paper we focus on the determination of the
trilinear composite Higgs self-coupling $\lambda_{HHH}$ in order to investigate
the prospects of testing the dynamics responsible for generating the
Higgs potential. This coupling is accessible in Higgs pair production
processes \cite{Djouadi:1999rca,Dawson:1998py}. In the SM the extraction of
$\lambda_{HHH}$ at the LHC is extremely challenging due to small cross
sections and large backgrounds
\cite{Lafaye:2000ec,Baur:2002rb,Baur:2003gpa}. Our goal is
to find parameter regions where $\lambda_{HHH}$ might be
accessible.\footnote{In Ref.~\cite{Barger:2003rs} a study of Higgs
  self-interactions was performed in the context of genuine dimension-six
  operators. Sensitivities were, however, derived for $e^+e^-$ linear
  colliders.}  We hope this to motivate realistic analyses taking into
account the relevant background processes and detector effects which
is beyond the scope of our paper.  \s

The organization of the paper is as follows.
In section \ref{sec:couplings} we will introduce the Higgs potential
and the general parametrization of the Higgs couplings. 
In section \ref{sec:hhcxn} the
various Higgs pair production cross sections at the LHC will be
analyzed. In section \ref{sec:sens} we will present and discuss
the sensitivity for the extraction of $\lambda_{HHH}$ before we
will conclude in section \ref{sec:concl}.

%%%%%%%%%%%%%%%%%%%%%%%%%%%%%%%%%%%%%%%%%%%%%%%%%%%%%%%
\section{Higgs potential and parametrization of the Higgs
  couplings} \label{sec:couplings}
The Holographic Higgs models of
Refs.~\cite{Contino:2003ve,Agashe:2004rs,Contino:2006qr} which are
based on a five-dimensional theory in Anti de-Sitter (AdS)
space-time, provide a resummation in the compositeness parameter
$\xi=(v/f)^2$. The bulk gauge symmetry $SO(5)\times U(1)_X \times SU(3)$
is broken down to the SM gauge group on the ultraviolet (UV) boundary and to
$SO(4)\times U(1)_X \times SU(3)$ on the infrared (IR). With the
symmetry-breaking pattern of the bulk and IR boundary given by $SO(5)
\to SO(4)$ we have four Goldstone bosons parametrized by the
$SO(5)/SO(4)$ coset. In this paper we will work in the framework of
the minimal models presented in
Refs.~\cite{Agashe:2004rs,Contino:2006qr}. With mild tuning, 
they are consistent with electroweak precision tests (EWPT). The
electroweak symmetry 
is broken dynamically via top loop effects. The Higgs Yukawa couplings
and hence the form of the Higgs potential of the low-energy effective
theory depends on the way the SM fermions are embedded in
representations of the bulk symmetry. In the minimal composite Higgs
model MCHM4
\cite{Agashe:2004rs} the SM fermions transform as spinorial
representations of $SO(5)$ whereas in the MCHM5 model
\cite{Contino:2006qr} as fundamental representations. The
Higgs potential is generated at one loop by gauge and fermion
interactions. In MCHM4, it is given by
\beq
V(h)= \alpha \cos\frac{h}{f} - \beta \sin^2
%_{\mbox{\scriptsize{MCHM4}}} 
\frac{h}{f} \;, 
\label{eq:VMCHM4}
\eeq
where $\alpha,\beta$ are integral functions of the form factors in the
low-energy effective Lagrangian encoding the effect of the strong
dynamics. EWSB is triggered by the top loops, and the Higgs
field $h$ acquires a vacuum expectation value (VEV) $v$ defined by
\beq
v \equiv f \sqrt{\xi} = f \sin \frac{<h>}{f} = 246 \mbox{ GeV} \;.
\eeq
Expanding the Higgs potential around the VEV we
can read off directly the trilinear and quartic Higgs self-couplings
from
\beq
V(H)= V(<h>)
+ \frac{1}{2} M_H^2 H^2 +
\frac{1}{6} \sqrt{1-\xi}\,\lambda^{\mbox{\scriptsize{SM}}}_{H^3} \,H^3 +
\frac{1}{24} \left(1-\frac{7}{3}\xi \right) 
\lambda^{\mbox{\scriptsize{SM}}}_{H^4}\, H^4 \;, 
\label{eq:V4}
\eeq
where we have introduced the SM trilinear and quartic Higgs couplings
\beq
\lambda^{\mbox{\scriptsize{SM}}}_{H^3} = \frac{3M_H^2}{v} \quad\quad
\mbox{and}
\quad \quad \lambda^{\mbox{\scriptsize{SM}}}_{H^4} = \frac{3M_H^2}{v^2} \;.
\eeq
The mass squared of the Higgs field fluctuation $H$ around the minimum is
given by
\beq
M_H^2 = \frac{4\beta^2-\alpha^2}{2\beta f^2} \;.
\eeq
The potential Eq.~(\ref{eq:VMCHM4}) can get further contributions from
additional heavy fields \cite{Agashe:2004rs}, but they will not change
the trilinear Higgs coupling. 
In MCHM5, the Higgs potential reads
\beq
V(h) = \sin^2 \frac{h}{f} \left( \alpha -\beta \cos^2
  \frac{h}{f} \right) \;.
\eeq
Expansion around the VEV leads to
\beq
V(H) = V(< h>) 
+ \frac{1}{2} M_H^2 H^2 + 
\frac{1}{6} \left(\frac{1-2\xi}{\sqrt{1-\xi}}\right)
\lambda^{\mbox{\scriptsize{SM}}}_{H^3} \,H^3 +
\frac{1}{24} \left(\frac{3-28\xi (1-\xi)}{3(1-\xi)}\right)
\lambda^{\mbox{\scriptsize{SM}}}_{H^4}\, H^4
\label{eq:V5}
\eeq
where
\beq
M_H^2 = \frac{2(\beta^2-\alpha^2)}{\beta f^2} \; .
\eeq
As can be inferred from Eqs.(\ref{eq:V4}) and (\ref{eq:V5}) the
trilinear and quartic Higgs couplings depend on the mass of the Higgs
boson and the parameter $\xi$. This is different from the SM, where they
are uniquely determined by the mass of the Higgs boson. The Higgs
couplings to fermions and gauge bosons are also 
modified compared to the SM couplings by corrections of the order
$\xi$. The modification factors of the interactions relevant for
our analysis are summarized in Table~1 for MCHM4 and
MCHM5. \s
\begin{table}
\begin{center}
$
\renewcommand{\arraystretch}{1.5}
\begin{array}{|c|c|c|c|c|} \hline
\mbox{Model/Coupling} & HVV & HHVV & Hff & HHH\\ \hline\hline
\mbox{MCHM4} & \sqrt{1-\xi} & 1-2\xi &
\sqrt{1-\xi} & \sqrt{1-\xi} \\ \hline
\mbox{MCHM5} & \sqrt{1-\xi} & 1-2\xi &
\frac{1-2\xi}{\sqrt{1-\xi}} & \frac{1-2\xi}{\sqrt{1-\xi}} \\ \hline
\end{array}
$
\caption{The Higgs couplings to massive gauge bosons $V=Z,W^\pm$, to
  fermions $f$ and the trilinear Higgs self-coupling with respect to the
  SM couplings in MCHM4 and MCHM5
  \cite{Contino:2010mh,Espinosa:2010vn}.}
\end{center}
\label{table:coup}
\end{table}

Whereas in both models the Higgs gauge couplings are always reduced
compared to the SM, in MCHM5 this is not the case for the trilinear
Higgs and the Higgs couplings to fermions. Near the SM limit for
low values of $\xi$ these couplings are reduced, with a stronger
reduction in MCHM5 than in MCHM4, but for large values of $\xi$ the
couplings in MCHM5 rise again and can become much larger than in the SM. 
The couplings can even vanish in this
model (for $\xi=0.5$). This will significantly affect all Higgs
production processes involving Higgs fermion couplings. 
Despite the vanishing Higgs couplings to
fermions the fermion masses are still created by the
Higgs mechanism, since the direct coupling to the VEV is
non-zero. \s

A novel coupling which is relevant for our
analysis is the direct coupling of two Higgs bosons to two
fermions. It is suppressed by another power of $v$ compared to the SM
Higgs Yukawa coupling and is given by 
\beq
\mbox{MCHM4:}&& \quad HHff: \; \xi \, \frac{m_f}{v^2} \\
\mbox{MCHM5:}&& \quad HHff: 4\xi \, \frac{m_f}{v^2} \;.
\eeq
In the SM limit, $\xi =0$, it vanishes as expected. \s

Before we turn to our analysis a comment on the constraints on
$\xi$ from direct searches at LEP and Tevatron as well from indirect
constraints due to electroweak precision measurements is at
order.\footnote{See Ref.~\cite{Espinosa:2010vn} for details.}
The direct search at LEP excludes $\xi \approx 0.7-0.95$ for $M_H$
ranging from $\sim 115$ to 80 GeV.  Low $\xi$ values are
excluded by the Tevatron search for $M_H \approx 162-167$ GeV. 
In MCHM5 an additional region for large
$\xi$ values ranging from $\sim 110-200$ GeV is excluded. \s

Concerning the EW precision data, in composite models there are three
main contributions to the oblique parameters \cite{Peskin:1991sw}. 
The contribution to the $T$ parameter would impose a very large
compositeness scale $f$. If we assume, however, the custodial symmetry to be
preserved by the strong sector, there is no contribution to the $T$
parameter. The models which we consider fulfill this requirement. The
contribution to the $S$ parameter imposes a lower bound on the mass of
the heavy resonances, $m_\rho \gsim 2.5$~TeV. We assume
the mass gap between the Higgs boson and the strong sector resonances
to be large enough to fulfill this constraint. Finally, since the
Higgs couplings to the SM gauge bosons are altered by corrections of the order
$\xi$ the cancellation between the Higgs and gauge boson contributions
to $S$ and $T$ does not hold anymore in contrast to the SM
case. They are hence both logarithmically divergent
\cite{Barbieri:2007bh}.\footnote{If the strong sector
  is invariant under custodial symmetry, the divergence in $T$ will be
  finally screened by the heavy resonances.} This leads to an upper
bound on the compositeness parameter of $\xi \lsim 0.2-0.4$ in the
Higgs mass range $M_H=80-200$ GeV \cite{Agashe:2005dk}. 
This bound can be relaxed by about a factor 2, if we allow for a
partial cancellation of the order of 50\% with contributions from
other states \cite{Espinosa:2010vn}. \s

Perturbative requirements forbid the limit $\xi \to 1$,
especially for MCHM5. A rough estimate shows that values close to 1
are allowed. The exact limit depends on the details of the models. 

%%%%%%%%%%%%%%%%%%%%%%%%%%%%%%%%%%%%%%%%%%%%%%%%%%%%%%%
\section{Higgs Pair Production Processes} \label{sec:hhcxn}
The Higgs potential is determined by the mass of the
physical Higgs boson and the trilinear and quartic Higgs
self-couplings. 
In the composite Higgs models the Higgs self-interactions are given by
$M_H$ and $\xi$, {\it i.e.} for $\lambda_{HHH}$,
\beq
\mbox{MCHM4:}&& \quad \lambda_{HHH} = \sqrt{1-\xi} \,\;
\lambda_{HHH}^{SM} \label{eq:self4}\\
\mbox{MCHM5:}&& \quad \lambda_{HHH} = \frac{(1-2\xi)}{\sqrt{1-\xi}}
\, \lambda_{HHH}^{SM} \label{eq:self5}\;,
\eeq
with the SM trilinear coupling being uniquely determined by $M_H$.
Through the measurement of $\lambda_{HHH}$ we make a first step
towards a full
reconstruction of the Higgs potential and gain insights in the
dynamics at the origin of EWSB. A departure from the SM relation
between Higgs boson mass and Higgs self-couplings would indicate New
Physics beyond the SM. \s

\begin{figure}[th]
\underline{double Higgs-strahlung: $q\bar{q}\to ZHH/WHH$} \\[0.2cm]
\vspace*{-0.6cm}
\begin{center}
\epsfig{figure=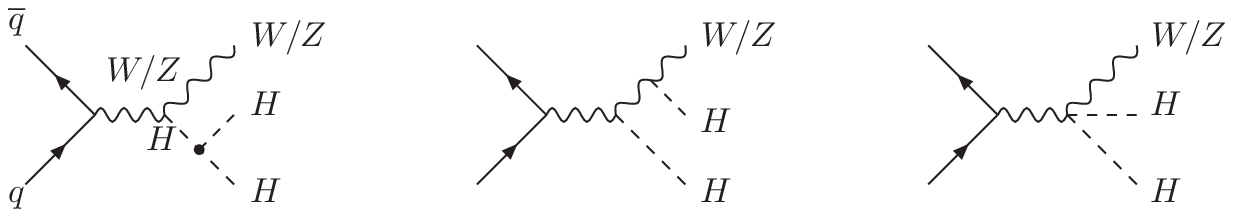,bbllx=110,bblly=625,bburx=600,bbury=685,width=16cm,clip=}
\end{center}
\underline{$WW/ZZ$ double Higgs fusion: $qq\to qqHH$} \\[0.2cm]
\vspace*{-0.5cm}
\begin{center}
\epsfig{figure=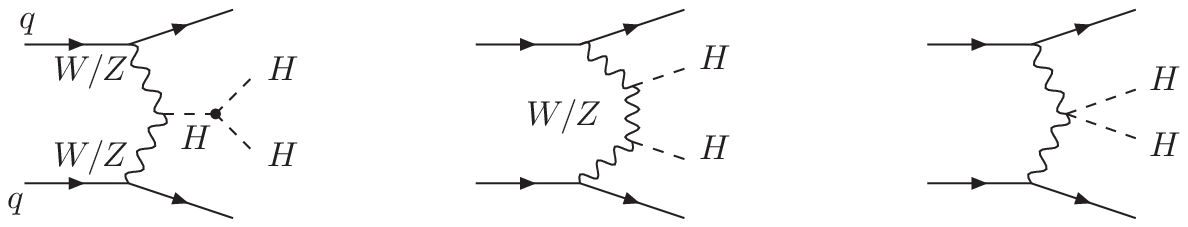,bbllx=110,bblly=620,bburx=600,bbury=685,width=16cm,clip=}
\end{center}
\underline{$gg$ double Higgs fusion: $gg\to HH$}
\begin{center}
\hspace*{-1.5cm}
\epsfig{figure=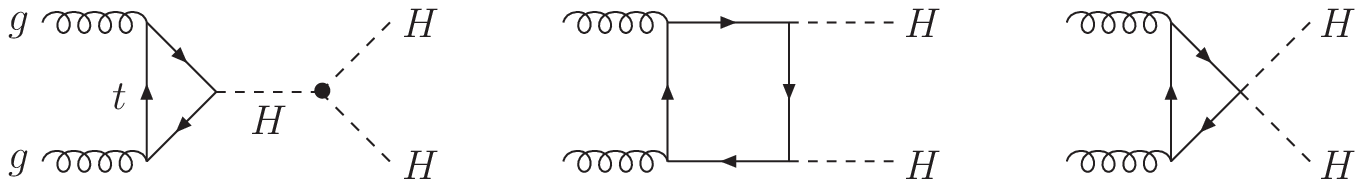,bbllx=110,bblly=615,bburx=600,bbury=685,width=14cm,clip=}
\caption{\label{fig:proddiags} Generic diagrams of the Higgs pair production
  processes at the LHC in the composite Higgs model: double Higgs-strahlung, $WW/ZZ$ fusion and $gg$ fusion.}
\end{center}
\end{figure}

The trilinear Higgs coupling can be measured directly in the
production of a pair of Higgs bosons. At the LHC, Higgs pairs can be
produced through double Higgs-strahlung off $W$ and $Z$ bosons
\cite{Barger:1988jk}, $WW$ and $ZZ$ fusion
\cite{Dobrovolskaya:1990kx}, and gluon gluon fusion \cite{Glover:1987nx}.
In principle, the cross sections in the composite Higgs model can
easily be derived from the SM cross sections by multiplying the SM Higgs
couplings with the corresponding modification factors,
cf. Table~1. There is one caveat, however. In gluon
fusion to a Higgs boson pair there is an additional diagram, 
which vanishes in the SM
limit and which involves the direct coupling of a pair of Higgs
bosons to two fermions. It is shown as the last diagram in 
$gg$ double Higgs fusion in Fig.~\ref{fig:proddiags}, which displays
the generic diagrams contributing to the Higgs pair production
processes at the LHC. 
The trilinear Higgs coupling is marked by a blob in the different processes. \s

The parton cross sections for double Higgs-strahlung $WHH$ and $ZHH$ 
have been
obtained from the corresponding results for
$e^+e^-$ collisions \cite{Djouadi:1999gv} by adjusting the couplings
properly and taking into account the modification of the composite
Higgs couplings with respect to the SM case. The gluon fusion cross
section has been derived from Ref.~\cite{Plehn:1996wb} by implementing the
appropriate correction factors of order $\xi$ in the Higgs
interactions and by adding the new diagram. This has been cross
checked in a second independent calculation. The production processes
at the LHC are then obtained by folding the double Higgs production
parton cross sections of the quark and gluon subprocesses,
respectively, $\hat\sigma (gg/qq' \to HH; \hat{s}=\tau s)$ with the
corresponding luminosities $d{\cal L}^{gg/qq'} / d\tau$,
\beq
\sigma (pp\to HH) = \int_{\tau_0}^1 d\tau \frac{d{\cal
    L}^{gg/qq'}}{d\tau} \hat{\sigma} (gg/qq' \to HH+X; \hat{s} = \tau s)
\eeq
with
\beq
\frac{d{\cal L}^{gg/qq'}}{d\tau} = \int_\tau^1 \frac{dx}{x} f^{g/q}(x;Q^2) \,
f^{g/q'}(\tau/x;Q^2)
\; ,
\eeq
where $f^{g/q^{(')}}$ denote the quark and gluon parton densities in
the proton at a typical scale $Q$. We have taken $Q=
\sqrt{\hat{s}}$ in gluon fusion, $Q= \sqrt{(M_V+2M_H)^2}$ in
Higgs-strahlung and $Q= M_V$ ($V=W,Z$) in the gauge boson
fusion processes. The kinematic threshold is denoted by $\tau_0$,
{\it i.e.} $\tau_0= 4M_H^2/s$ in gluon and gauge boson fusion and
$\tau_0=(M_V+2M_H)^2/s$ in Higgs-strahlung. The double gauge boson
fusion cross sections have been calculated with Madgraph/Madevent
\cite{Alwall:2007st} after implementing the composite Higgs model. \s

\begin{figure}[th]
\begin{center}
\epsfig{figure=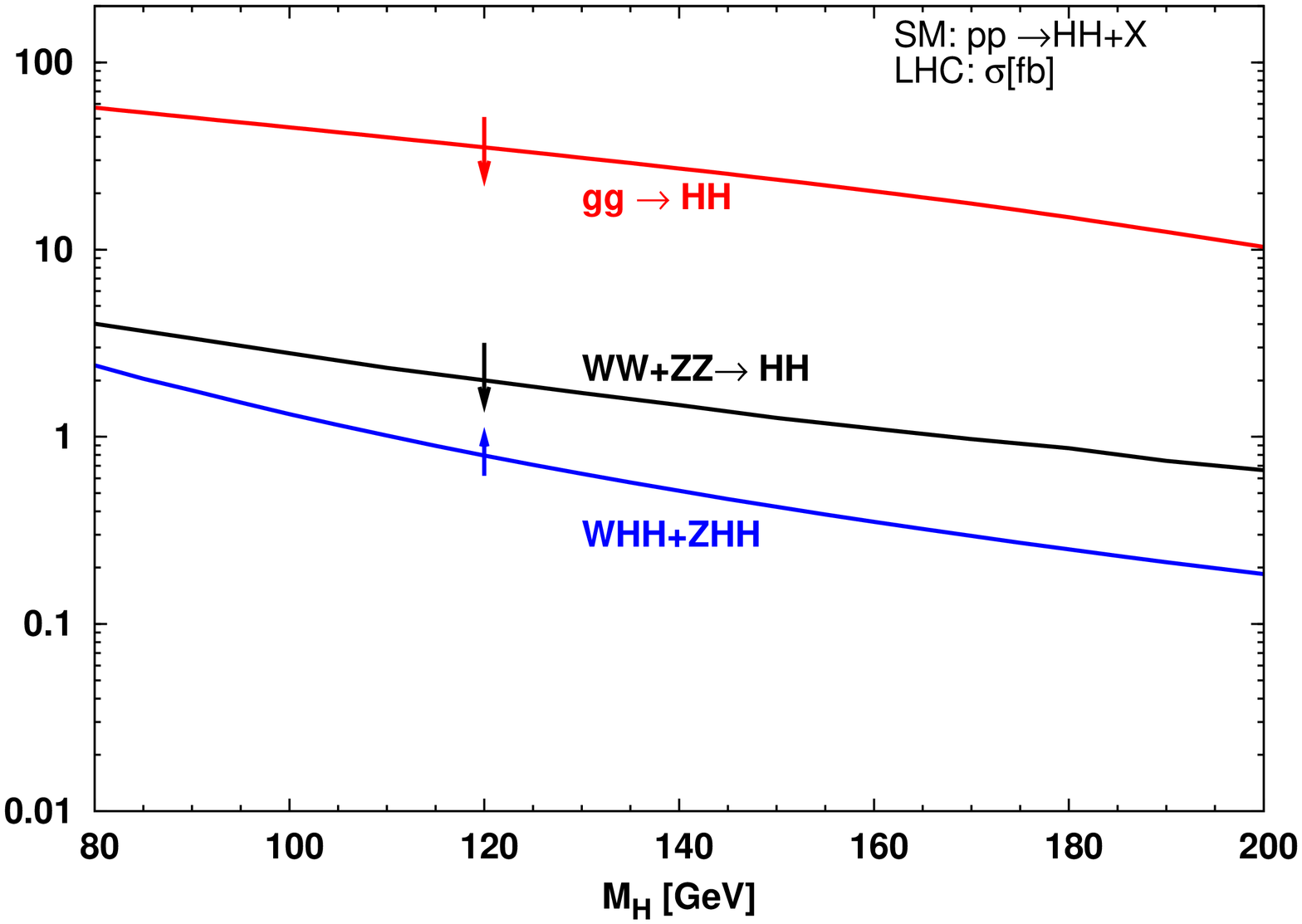,width=5cm,angle=-90} \hspace*{0.5cm}
\epsfig{figure=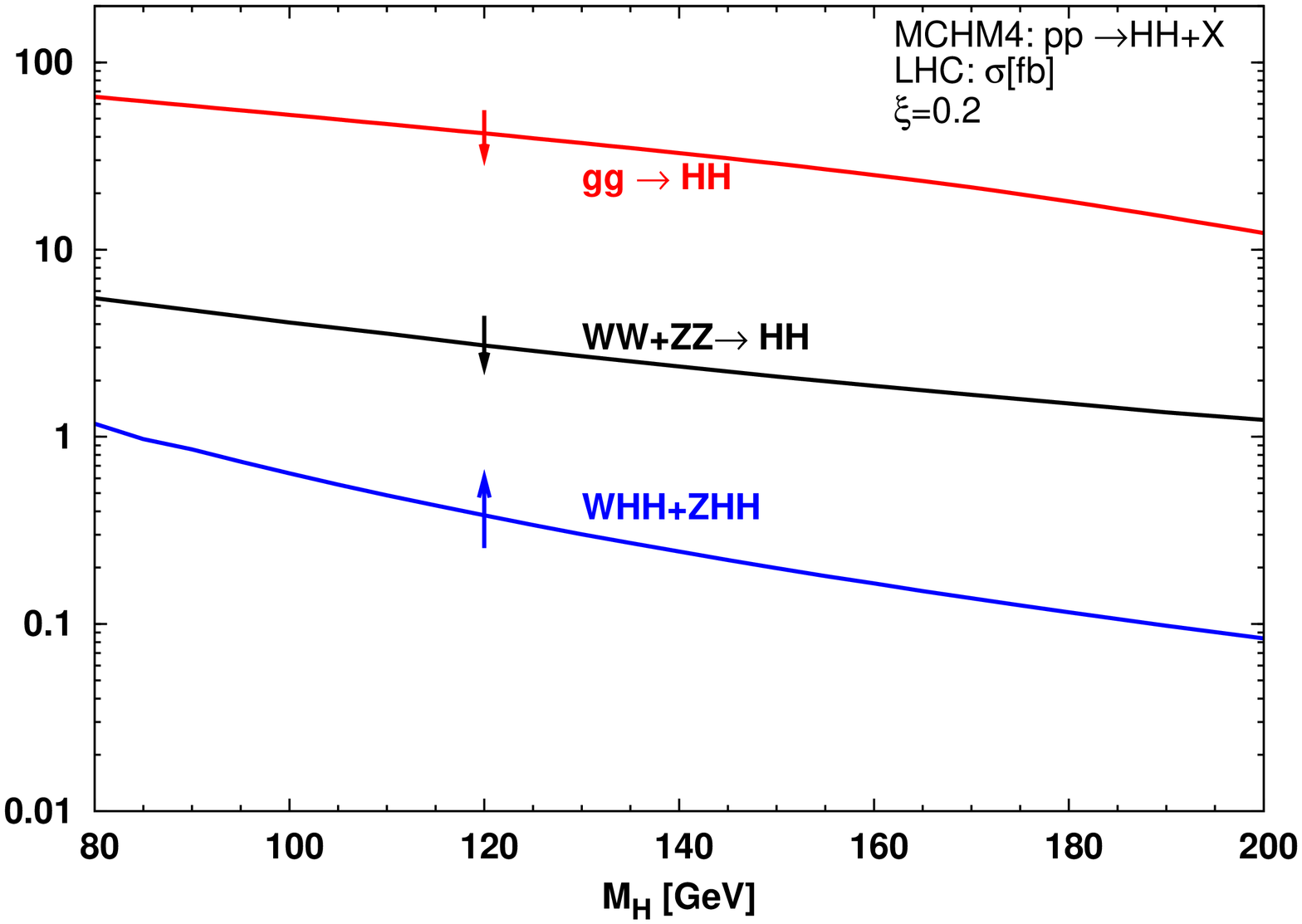,width=5cm,angle=-90} \\
\epsfig{figure=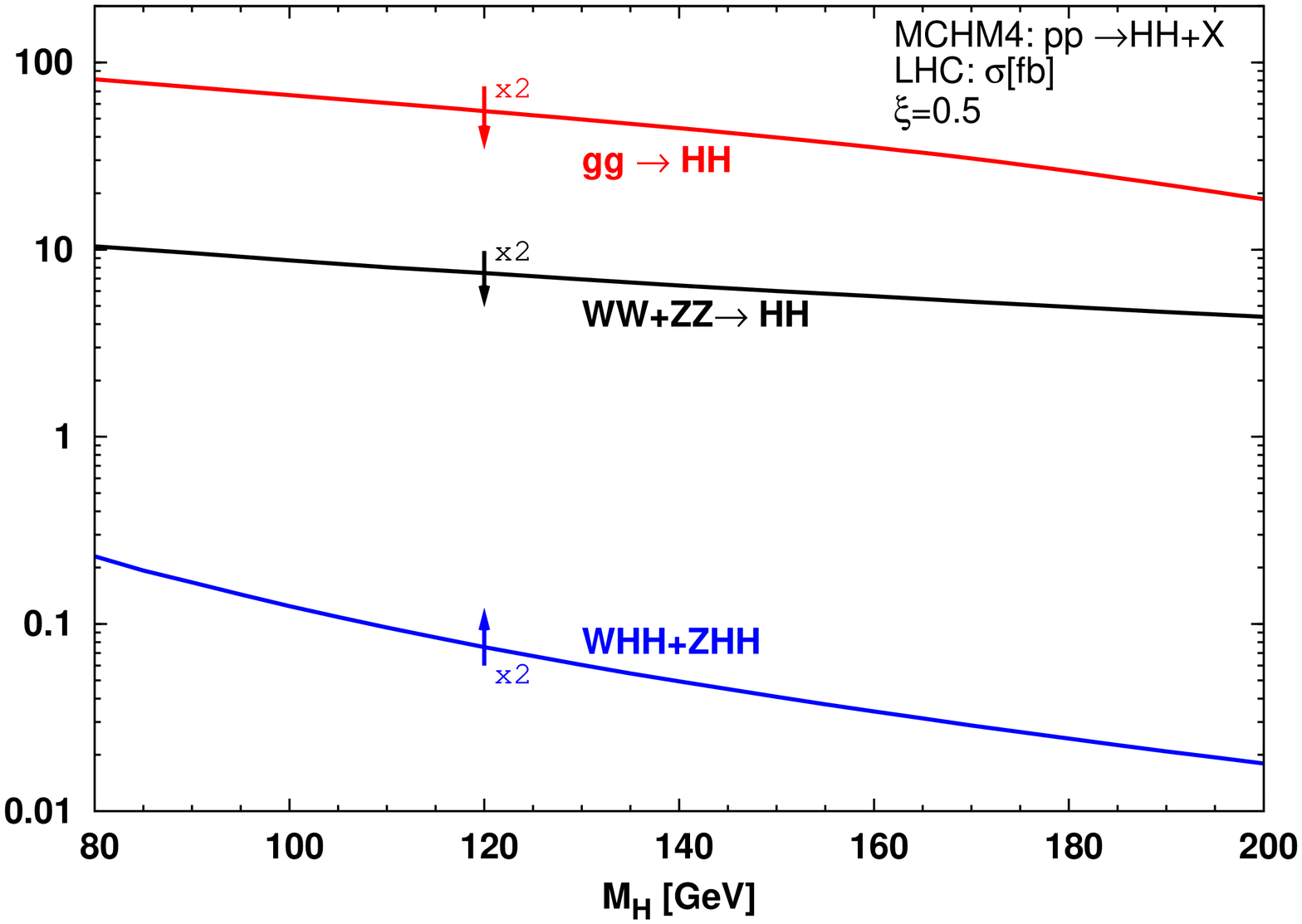,width=5cm,angle=-90} 
\hspace*{0.5cm}
\epsfig{figure=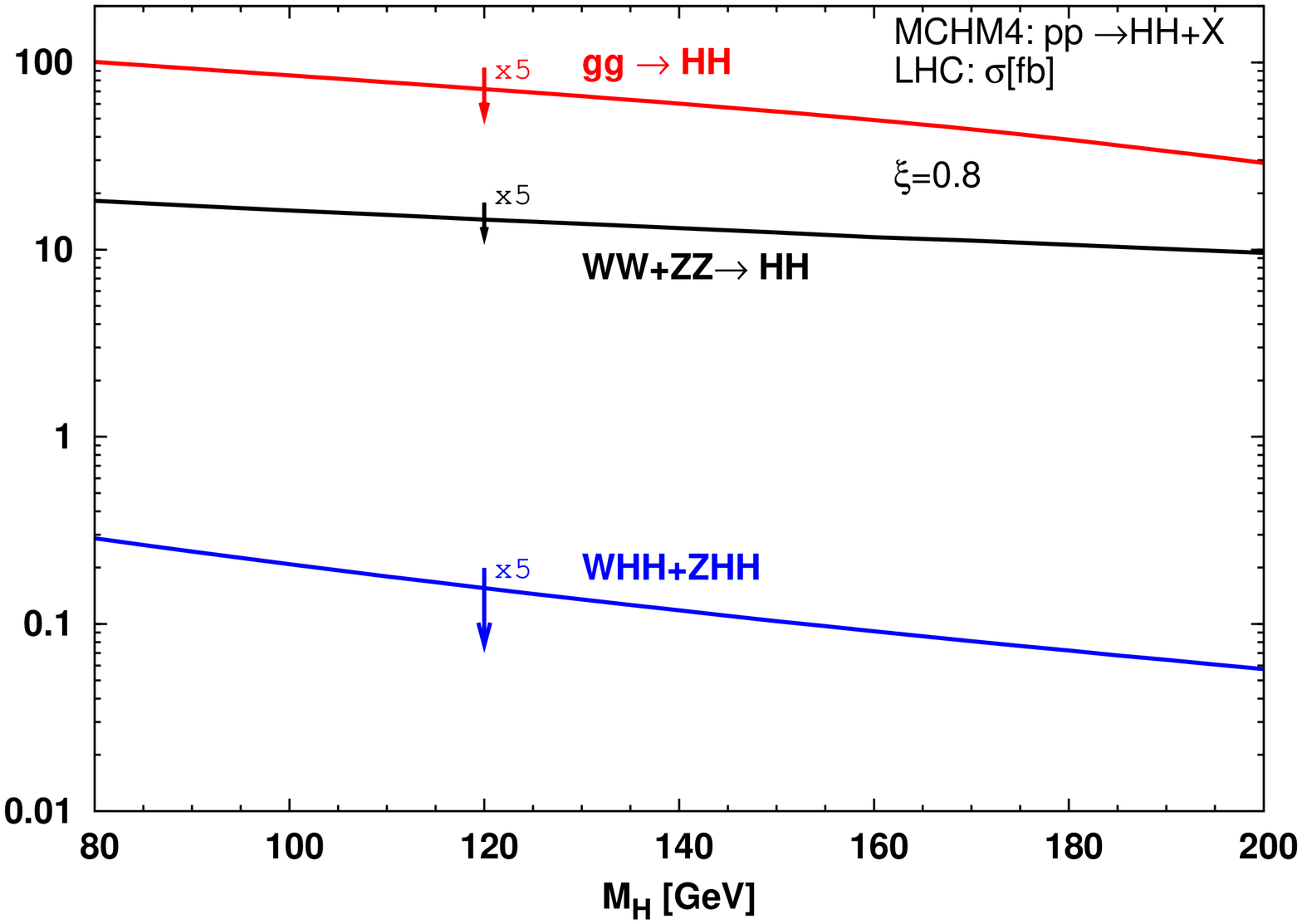,width=5cm,angle=-90} 
\caption{\label{fig:prodcxns4} Higgs pair production processes as a
  function of the Higgs boson mass in the SM ($\xi=0$, upper left) and
MCHM4 with $\xi = 0.2$ (upper right), 0.5 (bottom left) and 0.8
(bottom right). Arrows indicate the change in the cross section for a
variation of $\lambda_{HHH}$ from 0.5 to 1.5 its value in the
corresponding model.}
\end{center}
\vspace*{-0.4cm}
\end{figure}

\begin{figure}[th]
\begin{center}
\epsfig{figure=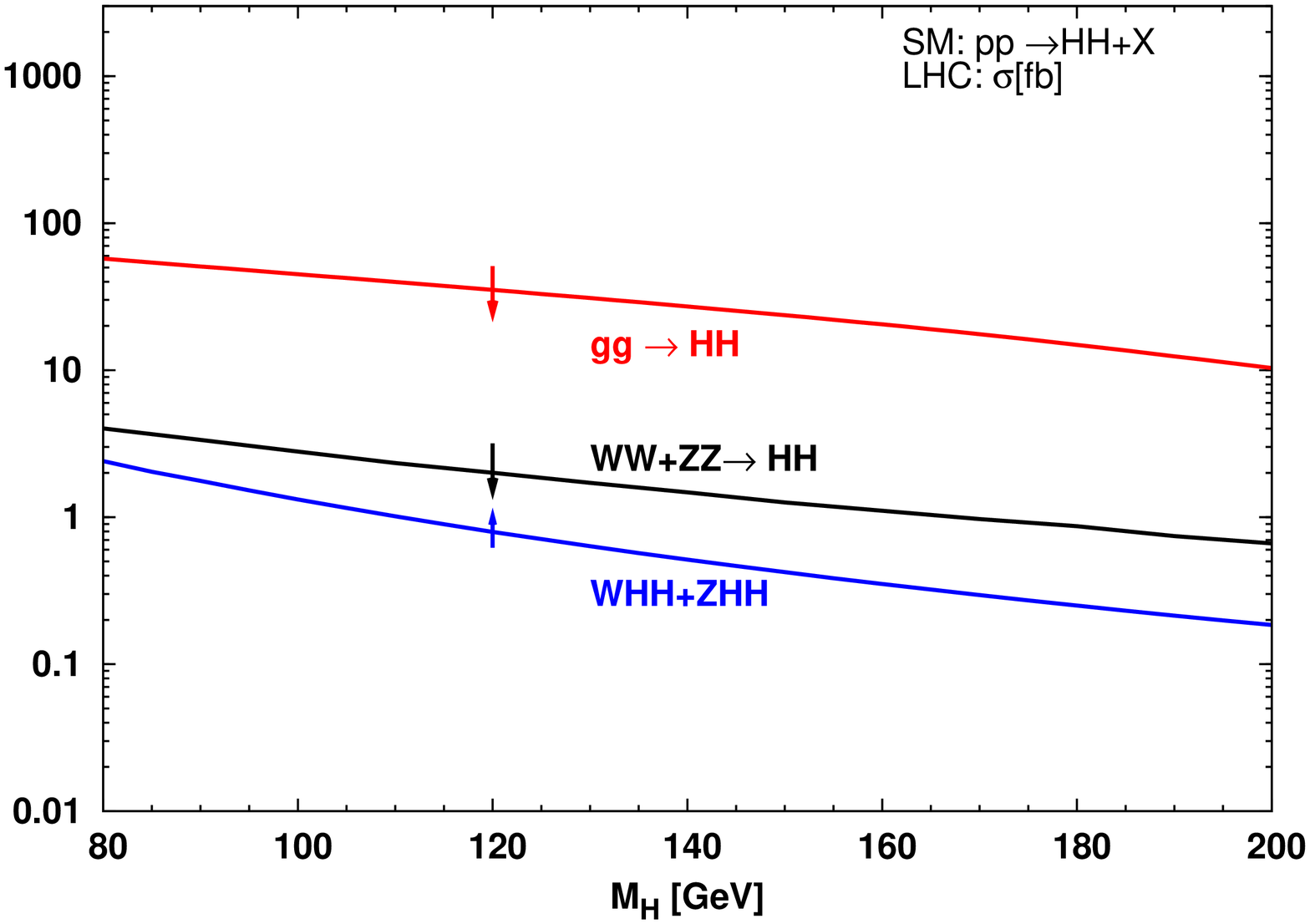,width=5cm,angle=-90} \hspace*{0.5cm}
\epsfig{figure=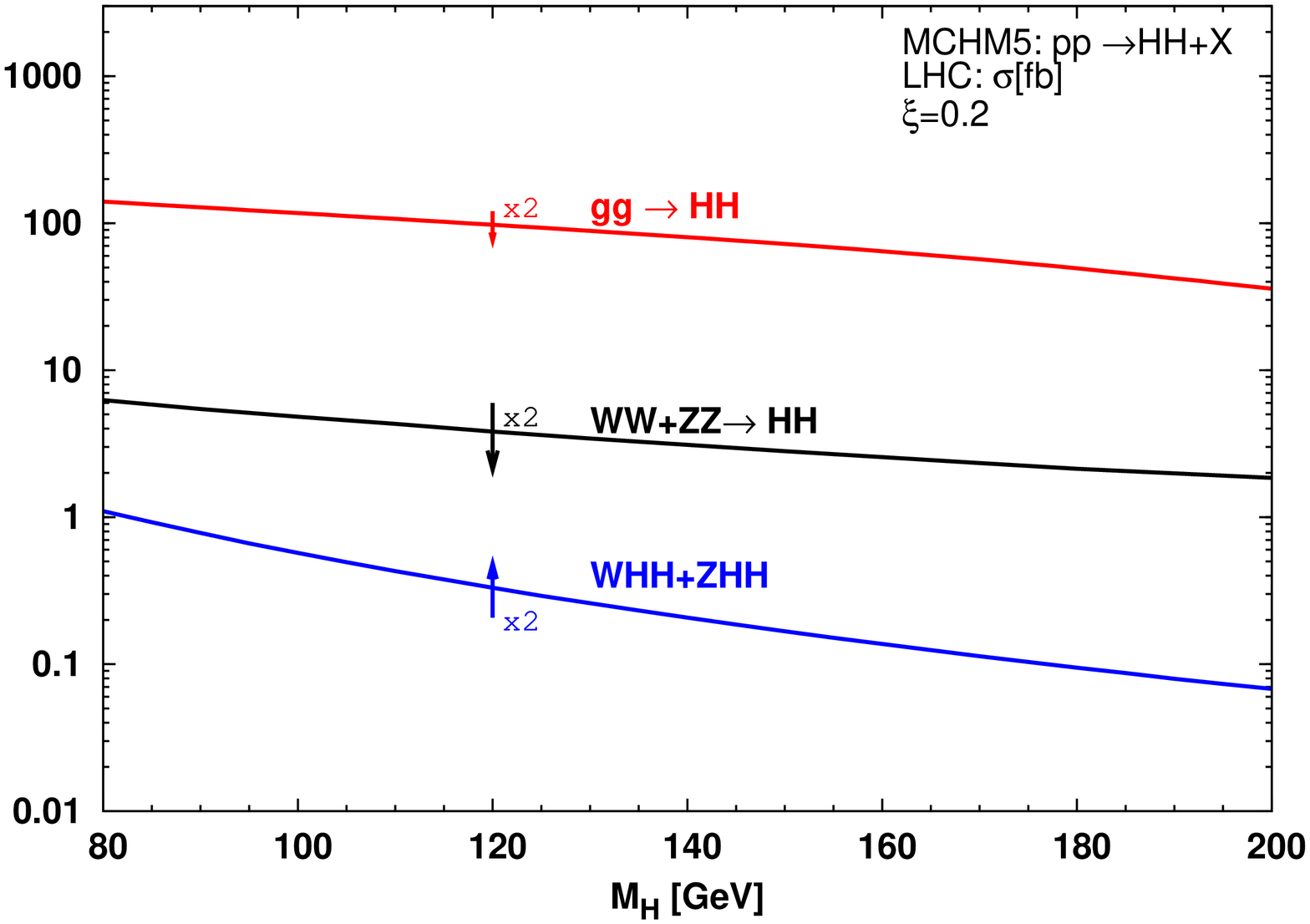,width=5cm,angle=-90} \\
\epsfig{figure=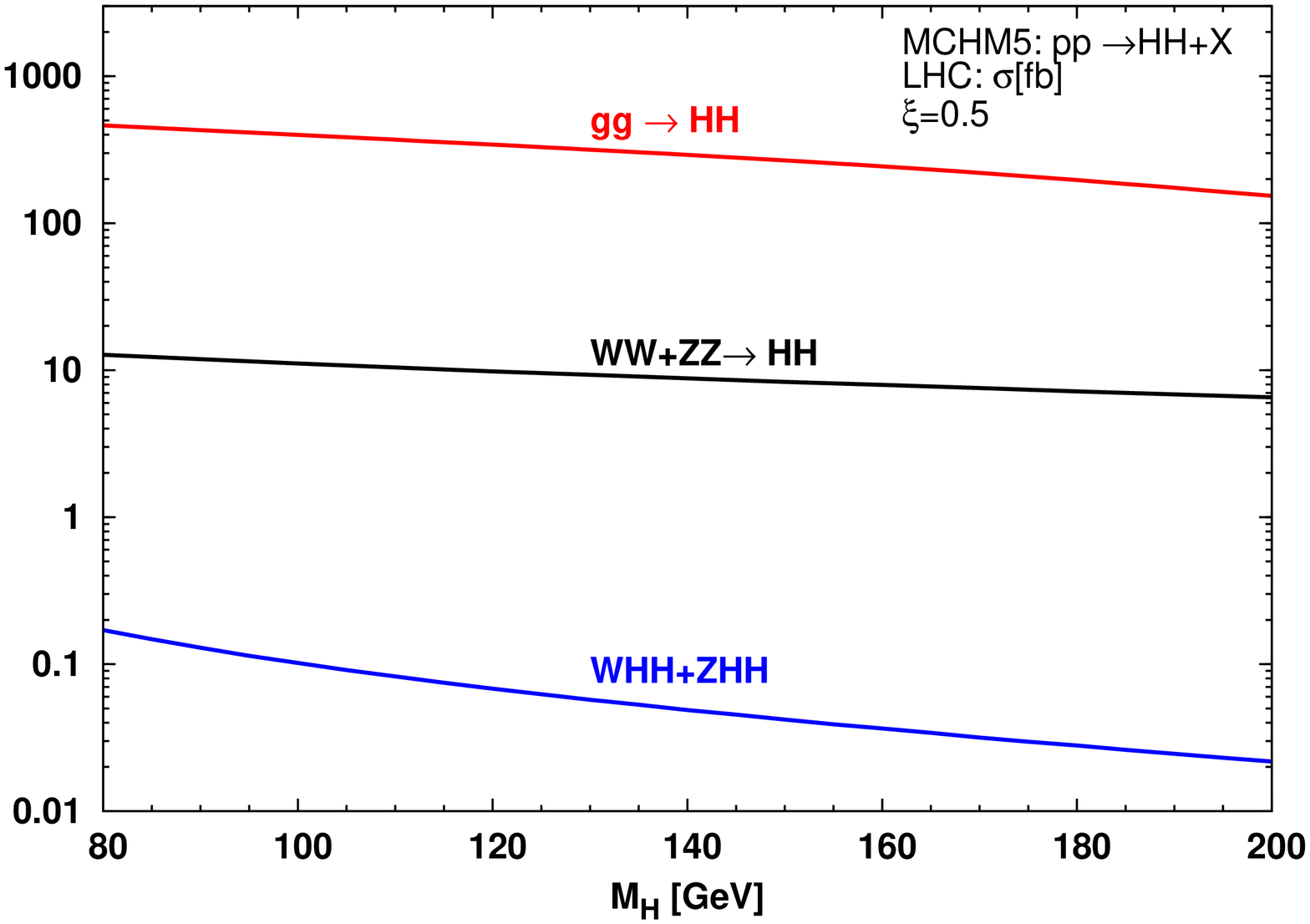,width=5cm,angle=-90} 
\hspace*{0.5cm}
\epsfig{figure=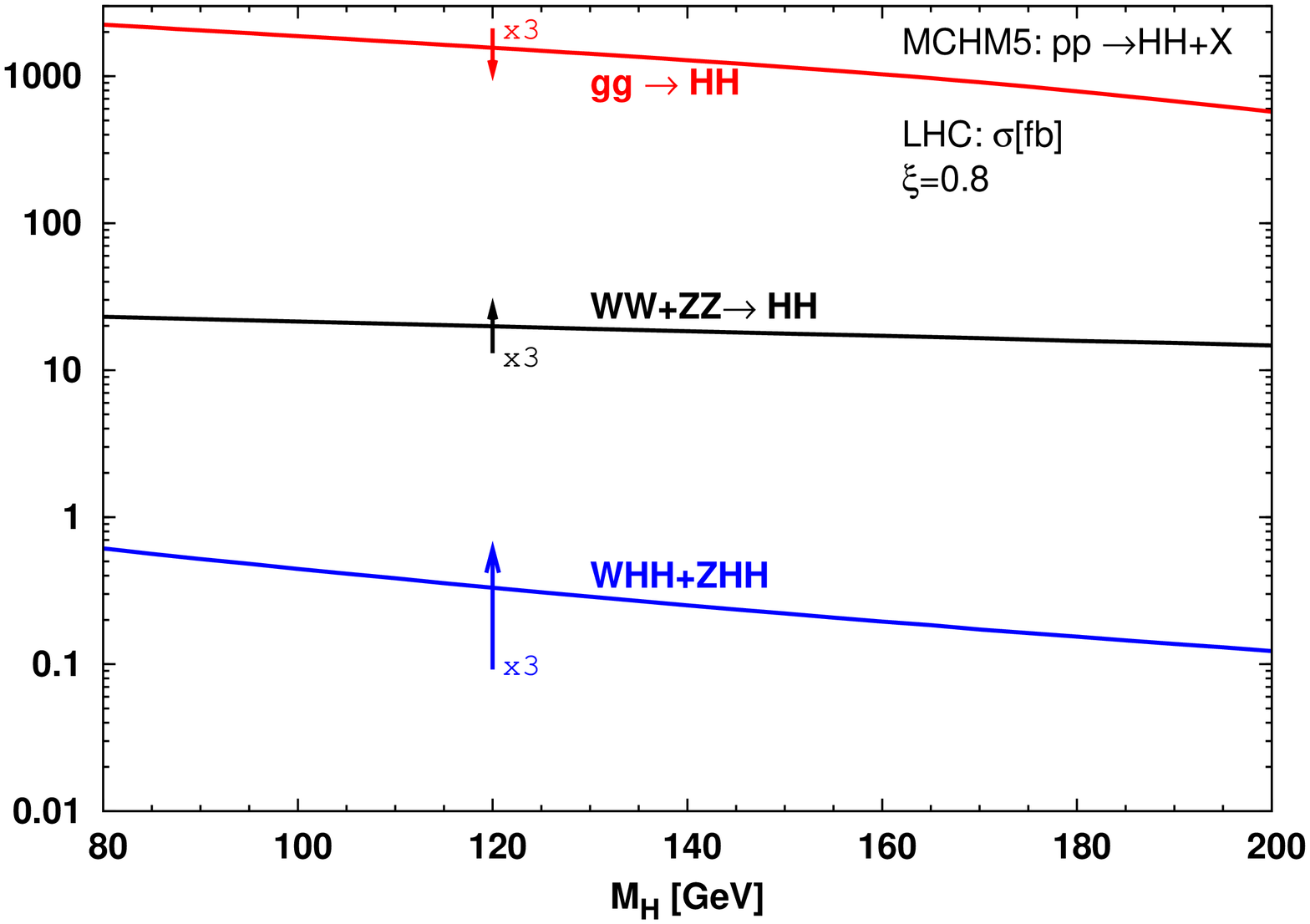,width=5cm,angle=-90} 
\caption{\label{fig:prodcxns5} The same as in Fig.\ref{fig:prodcxns4},
but for MCHM5.}
\vspace*{-0.4cm}
\end{center}
\end{figure}

Figure \ref{fig:prodcxns4} shows the double Higgs production processes as a
function of the Higgs mass $M_H=80-200$ GeV in the SM
and compared to the composite Higgs boson cross sections for MCHM4 and
three representative values of $\xi=0.2,0.5,0.8$. 
The corresponding plots for MCHM5 are displayed in
Fig.~\ref{fig:prodcxns5}. Throughout the whole paper we assume a
center-of-mass energy of 14 TeV, and we have adopted the parton
distribution functions CTEQ6L1 \cite{Pumplin:2002vw}. \s

%For gluon fusion the parton
%distribution function parametrizations CTEQ6.6AS \cite{Lai:2010nw}
%have been adopted, for gauge boson fusion and  double Higgs-strahlung 

As can be inferred from the figures, in the SM the gluon fusion
process is by far dominant due to the large number of gluons in
high-energy proton beams. In view of the results for single Higgs
production \cite{Spira:1995rr}, QCD radiative corrections are expected
to be important for this channel. In the low-energy limit for small
Higgs masses $M_H^2 \ll 4 M_t^2$ they lead to a $K$ factor of $K\approx
2$\cite{Dawson:1998py} in the mass range considered here. 
Since the QCD corrections of the SM can be translated in a
straightforward way to our composite Higgs model they have been
included in the plots. \s

The next important processes are given by the sum of $WW$ and $ZZ$
fusion, with the $WW$ fusion channel dominating over $ZZ$ fusion by a
factor $\sim 2.2-2.8$ for $\xi=0-0.8$. Double
Higgs-strahlung provides the smallest cross sections due to the
scaling behaviour $\sim 1/\hat{s}$. The cross sections for $WHH$ and
$ZHH$ have been summed with $WHH$ being larger by $\sim 1.6-2.1$ for
$\xi=0-0.8$. The vertical arrows in the figures show the change of the
cross sections for a variation of the Higgs self-coupling from 0.5 to
1.5 times its value in the corresponding model, {\it i.e.} with $\kappa \in
[0.5...1.5]$\footnote{The starting (end) point of the arrows corresponds to
  $\kappa=0.5$ (1.5).}
\beq
\begin{array}{lcll}
\lambda'_{HHH} &=& \kappa \lambda_{HHH}^{\mbox{\scriptsize SM}} &
\qquad \mbox{for the SM} \\[0.2cm]
\lambda'_{HHH} &=& \kappa \sqrt{1-\xi} 
\lambda_{HHH}^{\mbox{\scriptsize SM}} &
\qquad \mbox{for MCHM4} \\[0.2cm]
\lambda'_{HHH} &=& \kappa \frac{1-2\xi}{\sqrt{1-\xi}}
\lambda_{HHH}^{\mbox{\scriptsize SM}} &
\qquad \mbox{for MCHM5} \;.
\end{array}
\eeq
They indicate the sensitivity to $\lambda_{HHH}$ in the different models. Where
necessary amplification factors have been applied for the arrows to make them visible in the plots. \s

The size of the composite Higgs pair production cross sections with 
varying $\xi$ as well as their sensitivity to $\lambda_{HHH}$ can be
understood by examining the interference structure of the contributing
diagrams and the size and sign of the composite Higgs couplings. In
interpreting the gluon fusion cross section the additional new diagram
due to the direct Higgs pair coupling to $q\bar{q}$ has to be taken
into account. We start the discussion with \s

\noindent
\underline{$gg$ Fusion:} Due to the diagrams involving the
$HHq\bar{q}$ coupling the cross section increases with rising
$\xi$. The larger $HHq\bar{q}$ 
coupling in MCHM5 leads to a more important increase
than in MCHM4. The cross section can be up to a factor 30
bigger than in the SM with values of ${\cal O} (1 \mbox{
  pb})$. In MCHM4, the sensitivity gets smaller with rising $\xi$
due to the dominance of the $HHq\bar{q}$ diagram and the decreasing
Higgs self-coupling. This also holds for MCHM5 and $\xi < 0.5$.
The downwards orientation of the arrows is due to destructive interferences. 
In MCHM5, at $\xi=0.5$ the sensitivity completely vanishes since
$\lambda_{HHH}=0$.\footnote{A similar behaviour is found 
{\it e.g.} in the MSSM where the triple Higgs couplings vanish for
certain choices of the parameter space, {\it cf.}
Ref.~\cite{Djouadi:1999rca,Djouadi:1999gv}.} 
Although beyond $\xi=0.5$ the trilinear Higgs self-coupling in MCHM5
increases with rising $\xi$, the sensitivity does not, since it is
diluted by the more important contribution from the diagram with the
$HHq\bar{q}$ coupling, which does not suffer from a suppression due to
the Higgs boson propagator. \s

\noindent
\underline{$WW/ZZ$ Fusion:} The vector boson fusion cross section
increases with
rising $\xi$ but not as much as $gg$ fusion, so that 
with at most $\sim 20$ fb it reaches about 5 times the SM cross section.
For $\xi \le 0.5$ the rise is due to the destructive interference of
the diagram involving the $VVHH$ coupling with the $u$- and $t$-
channel diagrams, which gets smaller with the decreasing Higgs gauge
couplings in both models. Above $\xi=0.5$ the $VVHH$ coupling  
changes sign and
the interference becomes constructive leading to larger cross
sections. Since in MCHM5 also the $\lambda_{HHH}$ coupling changes
sign above $\xi=0.5$, the related previously destructive interference becomes
constructive and the arrow changes its orientation. \s

The suppression of the sensitivity to the trilinear coupling with
rising $\xi$ is due to the increasing dominance of the strong sector:
In the scattering amplitude of the longitudinal gauge bosons the
modified Higgs couplings to gauge bosons lead to an incomplete 
cancellation of the diagrams, which only involve Higgs gauge
couplings and which grow with the c.m. energy $s$. Thus in composite
Higgs models double Higgs production in $W_LW_L$ 
fusion becomes strong \cite{Giudice:2007fh ,Contino:2010mh}. This
behaviour can be inferred from the explicit formula of the amplitude
for longitudinal gauge boson scattering in a pair of Higgs bosons,
\beq
\mathcal{M} &=& G_{\scriptscriptstyle{F}}\frac{s}{\sqrt{2}}\Big\{ 
(1+\beta_{\scriptscriptstyle{W}}^2)
\left[(1-2\xi)+
  \frac{A\cdot\lambda^{\mbox{\scriptsize{SM}}}_{\scriptscriptstyle{HHH}}}{(s-M_{\scriptscriptstyle{H}}^2)/M_{\scriptscriptstyle{Z}}^2}\right]
\nonumber \\
&& +\frac{1-\xi}{\beta_{\scriptscriptstyle{W}}\beta_{\scriptscriptstyle{H}}}
\left[\frac{(1-\beta_{\scriptscriptstyle{W}}^4)+(\beta_{\scriptscriptstyle{W}}-\beta_{\scriptscriptstyle{H}}\cos\theta)^2}{\cos\theta
    -x_{\scriptscriptstyle{W}}} -	
\frac{(1-\beta_{\scriptscriptstyle{W}}^4)+(\beta_{\scriptscriptstyle{W}}+\beta_{\scriptscriptstyle{H}}\cos\theta)^2}{\cos\theta
  +x_{\scriptscriptstyle{W}}} \right] \Big\} \nonumber \\
&\stackrel{s\to\infty}{\longrightarrow} & -  \sqrt{2}
  G_{\scriptscriptstyle{F}} s \, \xi\;,
\eeq
with $\beta_{W,H}$ denoting the $W,H$ velocities and 
$\theta$ the Higgs production angle in the c.m. frame of $WW$, and
$x_W=(1-2M_H^2/s)/(\beta_W \beta_H)$. We have $A=1-\xi$ in MCHM4
and $A=1-2\xi$ in MCHM5. The high-energy limit agrees exactly with
the behaviour of the longitudinal gauge boson scattering 
amplitude given in Refs.~\cite{Giudice:2007fh ,Contino:2010mh}. 
\s

Both in $gg$ and $WW$ fusion for high energies the diagrams
without $\lambda_{HHH}$ will hence completely dominate and dilute the
sensitivity to $\lambda_{HHH}$. The distribution in the 
invariant mass of the Higgs pair, which is sensitive to
$\lambda_{HHH}$ \cite{Lafaye:2000ec,Baur:2002rb,Baur:2003gpa}, can
then only be exploited in the region close to the threshold to get access to the
trilinear Higgs coupling \cite{Contino:2010mh}.\footnote{We thank  Christophe Grojean for drawing our attention to this point.}
\s

\noindent
\underline{$WHH/ZHH$:} In the SM all diagrams interfere constructively.
For non-vanishing $\xi < 0.5$ all composite
Higgs couplings are suppressed compared to the SM so that the cross
sections in MCHM4 and 5 are less important. For $\xi > 0.5$, the
$HHVV$ coupling changes sign, and in MCHM4 the diagram interferes
destructively with the one involving $\lambda_{HHH}$. The
  cross section increases with rising $\xi$, {\it i.e.}  smaller Higgs
  self-coupling, and the arrow changes
  orientation. In MCHM5, also the trilinear Higgs coupling changes
  sign and the interference between the two diagrams remains constructive.
For small $\xi$ values the sensitivity is larger in MCHM4, since
$\lambda_{HHH}$ is bigger than in MCHM5. For large values of $\xi$,
$\lambda_{HHH}$ is larger in MCHM5 and thus the sensitivity, too.
The cross sections are small and do hardly exceed 1 fb in the
composite Higgs models. 

%%%%%%%%%%%%%%%%%%%%%%%%%%%%%%%%%%%%%%%%%%%%%%%%%%%%%%%

\section{Sensitivities} \label{sec:sens}
 
\begin{figure}[ht]
\begin{center}
\epsfig{figure=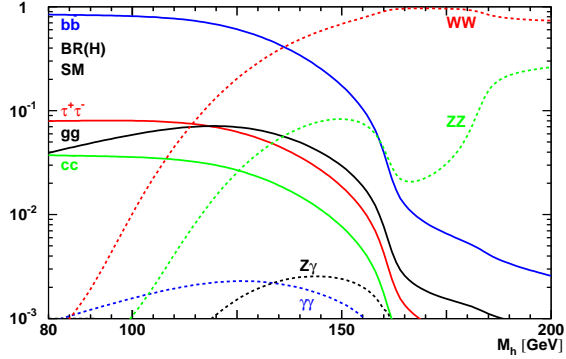,height=5cm}
\caption{The branching ratios in the SM and MCHM4 as a function of $M_H$. }
\label{fig:smbranch}
\end{center}
\vspace*{-0.2cm}
\end{figure}

\begin{figure}[ht]
\begin{center}
\epsfig{figure=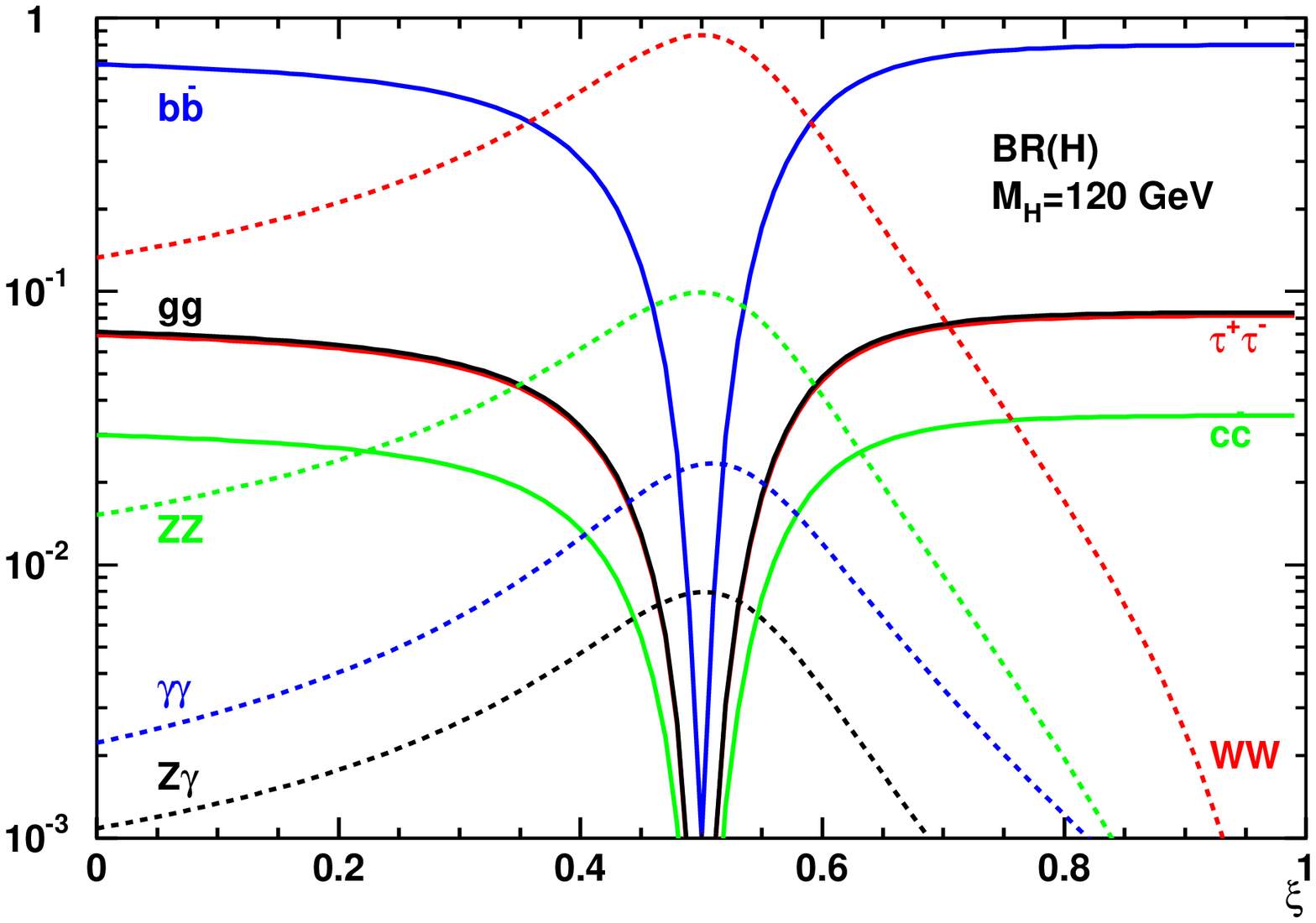,height=5cm}
\epsfig{figure=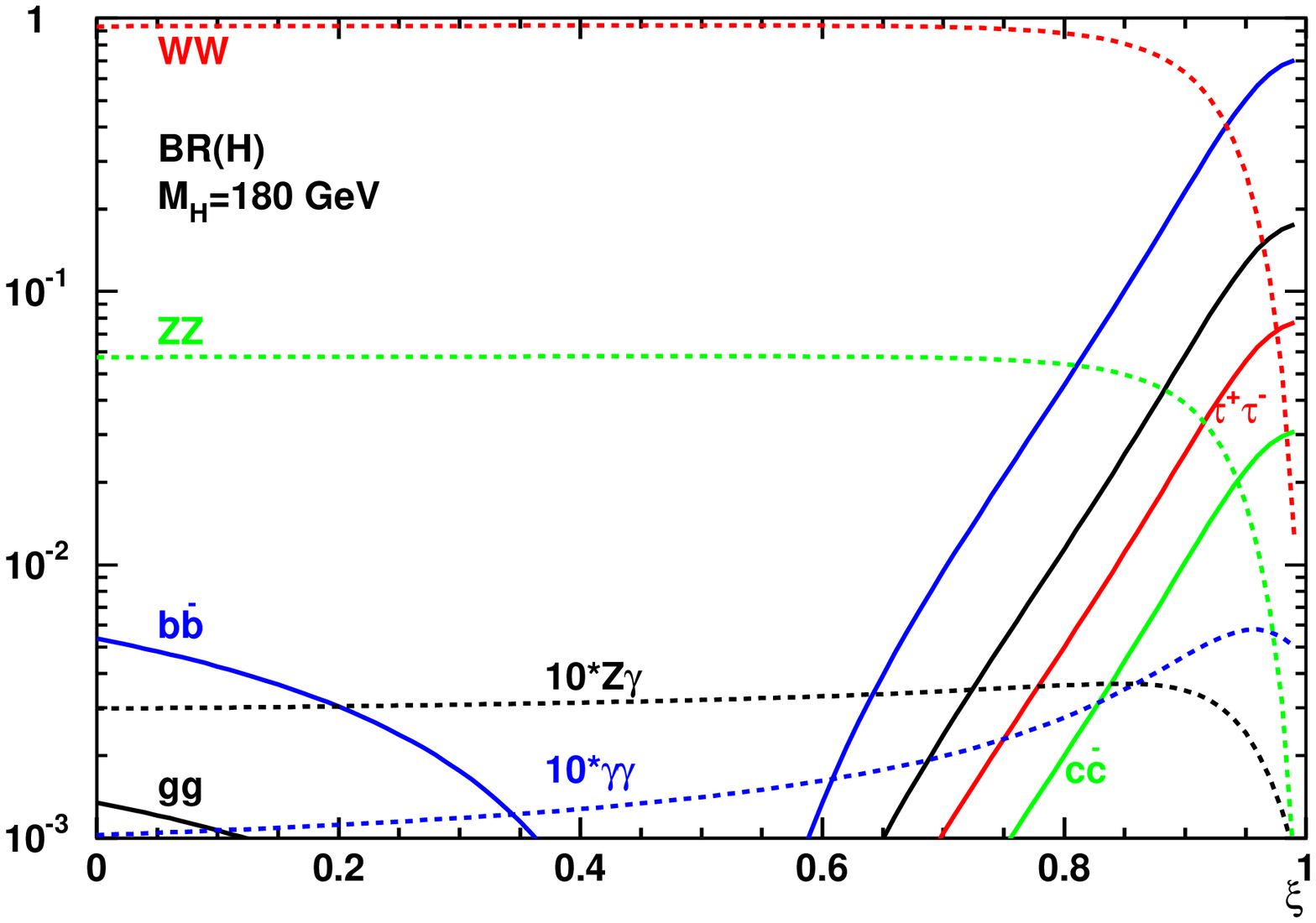,height=5cm}
\caption{The branching ratios in MCHM5 as a function of $\xi$ for
  $M_H=120$~GeV (left) and $M_H=180$~GeV (right) \cite{Espinosa:2010vn}.}
\label{fig:branch}
\end{center}
\vspace*{-0.4cm}
\end{figure}

In this section we will discuss the sensitivity of the Higgs pair
production cross sections to the trilinear Higgs
coupling. We will ask two questions: 
\begin{itemize}
\item[1.)] What are the prospects to distinguish composite Higgs pair
  production from the SM case?
\item[2.)] What are the prospects to extract the trilinear composite Higgs
  coupling from the double Higgs production processes?
\end{itemize}
We will investigate these questions by constructing sensitivity areas for
various final states. The final states are dictated by the Higgs branching ratios.
These depend on the mass of the Higgs boson and the value of
$\xi$. Since in MCHM4 the Higgs couplings to a pair of gauge bosons and
fermions are modified by the same factor, the branching ratios
do not change with respect to the SM
\cite{Contino:2010mh,Espinosa:2010vn}. These are shown as a function
of the Higgs boson mass in Fig.~\ref{fig:smbranch}. They have been
obtained by means of the program HDECAY \cite{hdecay}.
As can be inferred
from the figure,  for $M_H$ below the gauge boson threshold the Higgs
boson dominantly decays into $b\bar{b}$. Above $\sim 140$ GeV the main
decay channel is given by the $W^+W^-$ final state, followed by $ZZ$. 
Figures~\ref{fig:branch} show the branching ratios in MCHM5 as
function of $\xi$ for two values of $M_H=120$ and 180 GeV.\footnote{
 For the generation of the branching ratios the Higgs coupling
 modifications in MCHM5 have been implemented in the program
 HDECAY \cite{hdecay}.} The SM limit corresponds to $\xi=0$ in
Figs.~\ref{fig:branch}. In MCHM5, the behaviour can be drastically
different from the SM. For Higgs masses below the gauge boson
threshold and small values of $\xi$, the Higgs boson dominantly decays
into $b\bar{b}$, as in the SM. For $\xi=0.5$, however, where the
coupling to fermions vanishes\footnote{The Higgs boson decay
  into fermions through electroweak particle-loops cannot compete with
  the loop-mediated $\gamma\gamma$ decay, since it has in addition to the
  loop suppression factor a suppression factor of order $m_f^2 /M^2$ where
  $m_f$ is the light fermion mass and $M$ is a mass of electroweak size
  that can be either the Higgs mass, the top mass or the W mass
  depending on the diagram involved. The decay is about 2 orders of magnitude
  subdominant compared to the $\gamma\gamma$ decay
  \cite{Espinosa:2010vn}.}, the decays into gauge bosons dominate
and the rare decay into a clean final state photon pair can be as large
as a few percent. Above the gauge boson threshold the decays are
dominated by $W^+W^-$ and $ZZ$ final states up to near the
technicolor limit, where the Higgs gauge couplings are very small and
the Higgs fermion couplings are enhanced, so that the decay into
$b\bar{b}$ is the most important one.

For the production processes we concentrate on $gg$ fusion. In
gauge boson fusion the additional two tagging jets
could be exploited. However, the cross section is an order of magnitude
smaller than $gg$ fusion. The Higgs radiation cross sections, too, are too small
to be accessible. The smallness of the Higgs pair production cross section
and the large backgrounds 
represent a considerable challenge. For Higgs boson masses below the
gauge boson threshold the decay into $b\bar{b}$ has to be combined
with a rare decay of the second Higgs boson, since the 4$b$ final state
is hopeless in view of the huge QCD background. We therefore
investigate the $b\bar{b}\gamma\gamma$ and the $b\bar{b}\tau\tau$ 
final states. For Higgs masses above $\sim 140$
GeV, the $4W$ final state is most promising. We first address the question: \s

\noindent
\underline{1.) Can composite Higgs pair production be distinguished
  from the SM case?} 

\begin{figure}[ht]
\begin{center}
\epsfig{figure=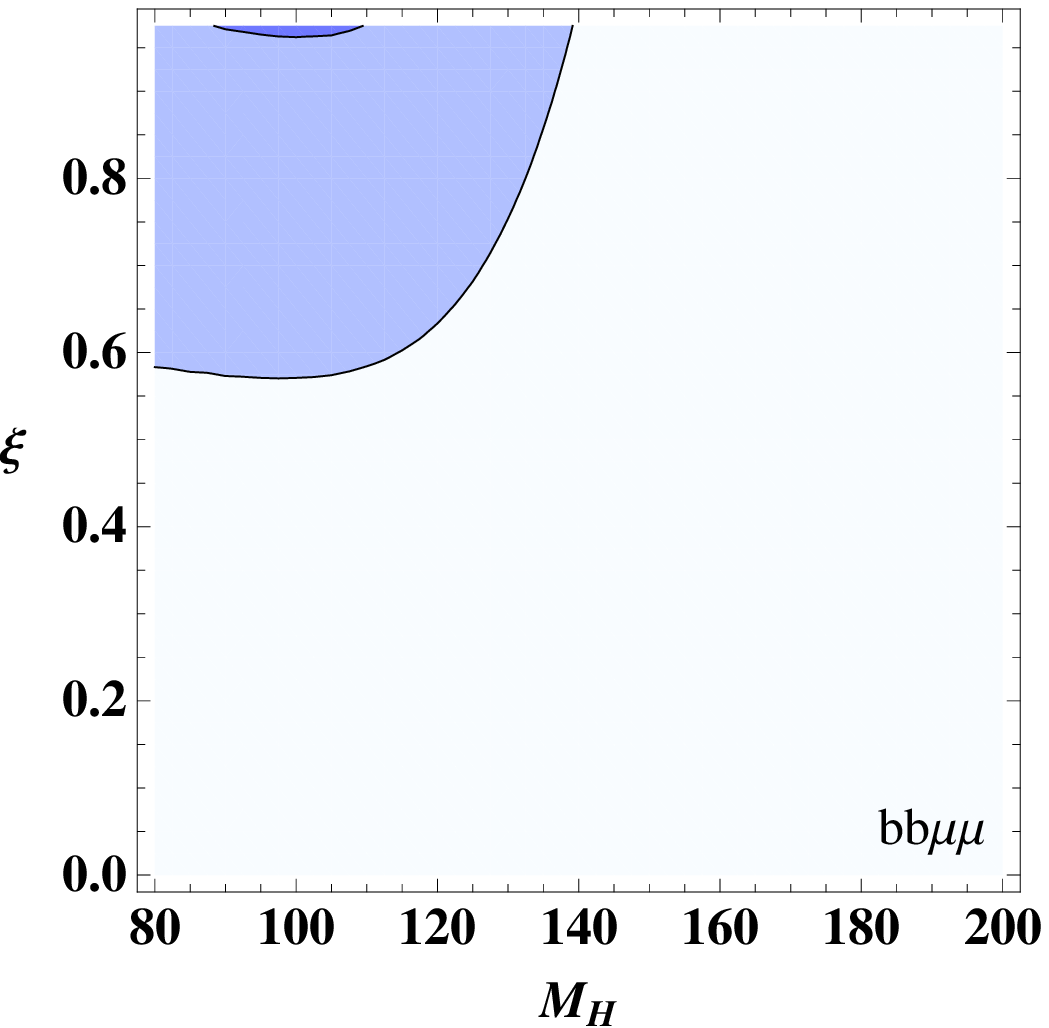,height=3.8cm}
\epsfig{figure=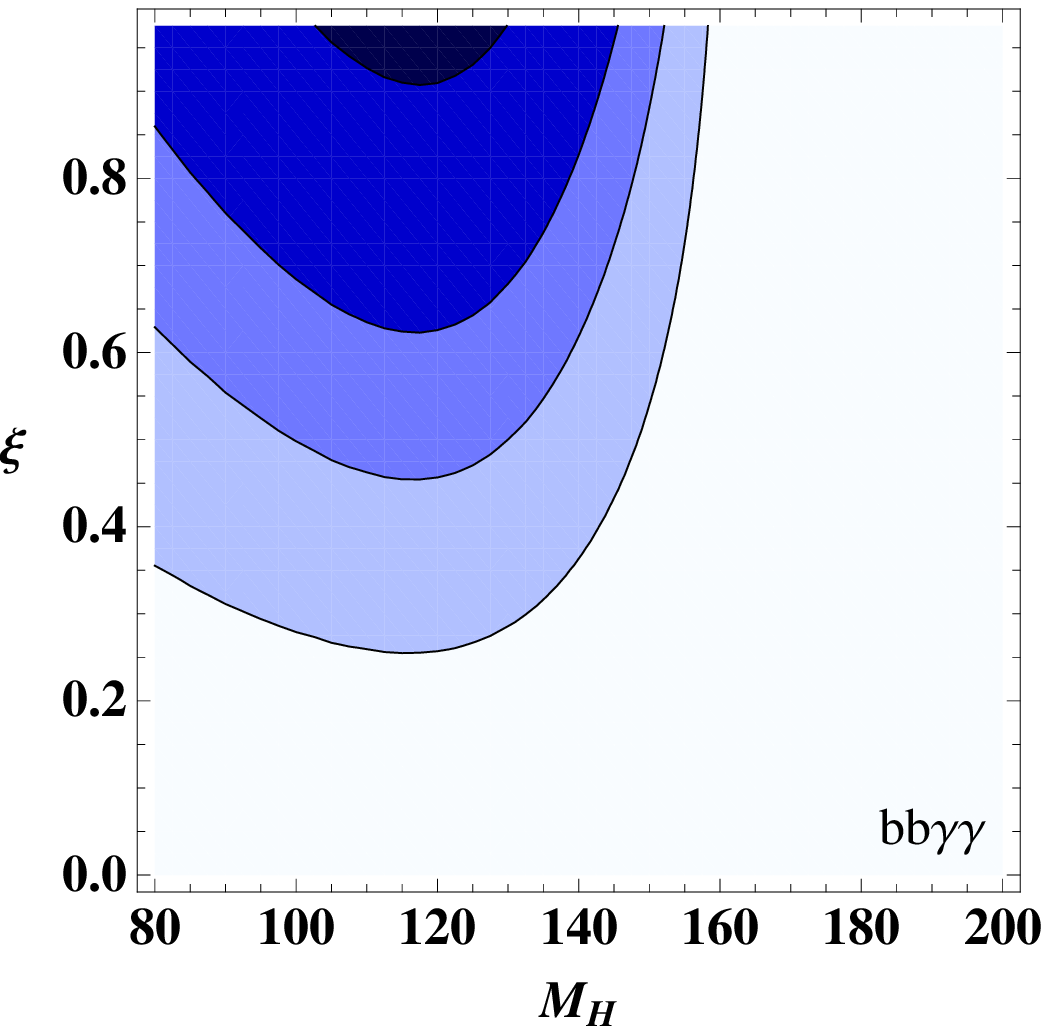,height=3.8cm}
\epsfig{figure=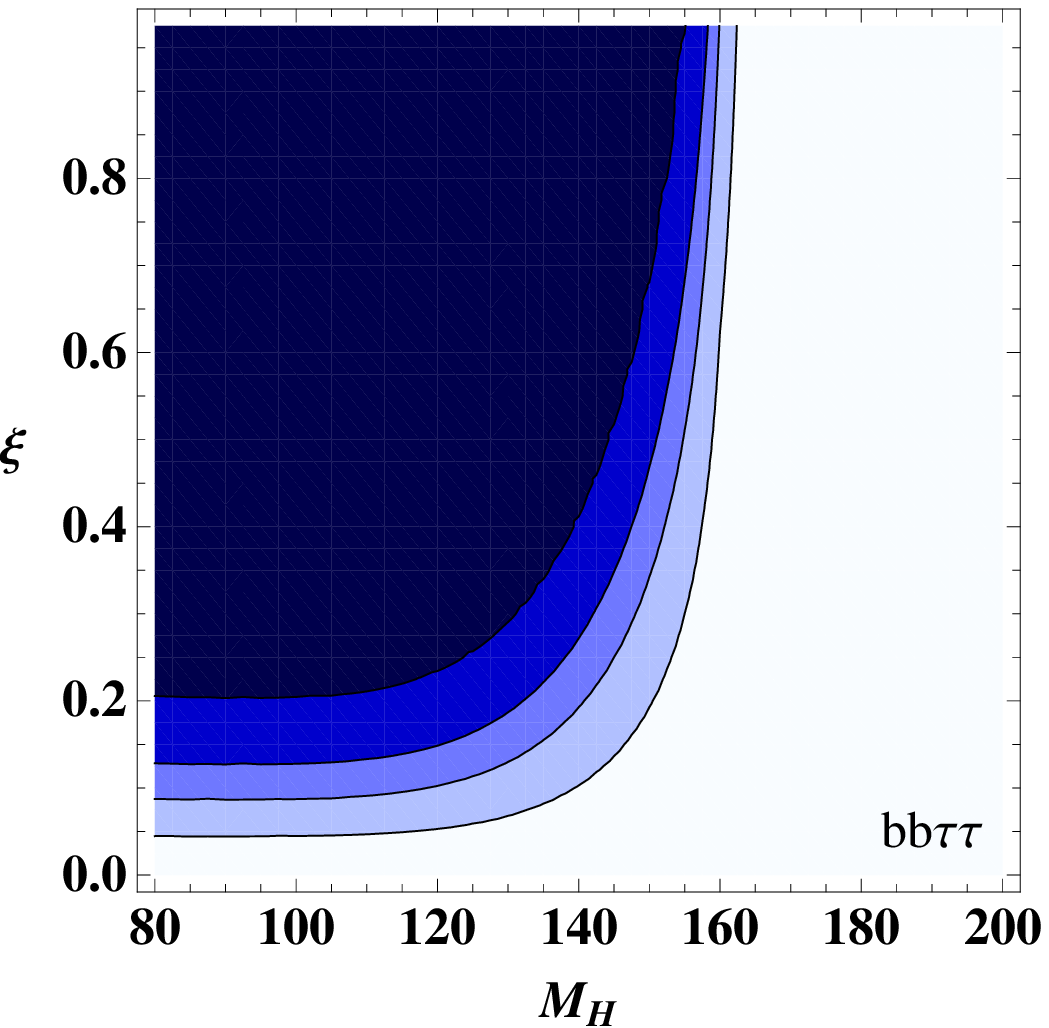,height=3.8cm}
\epsfig{figure=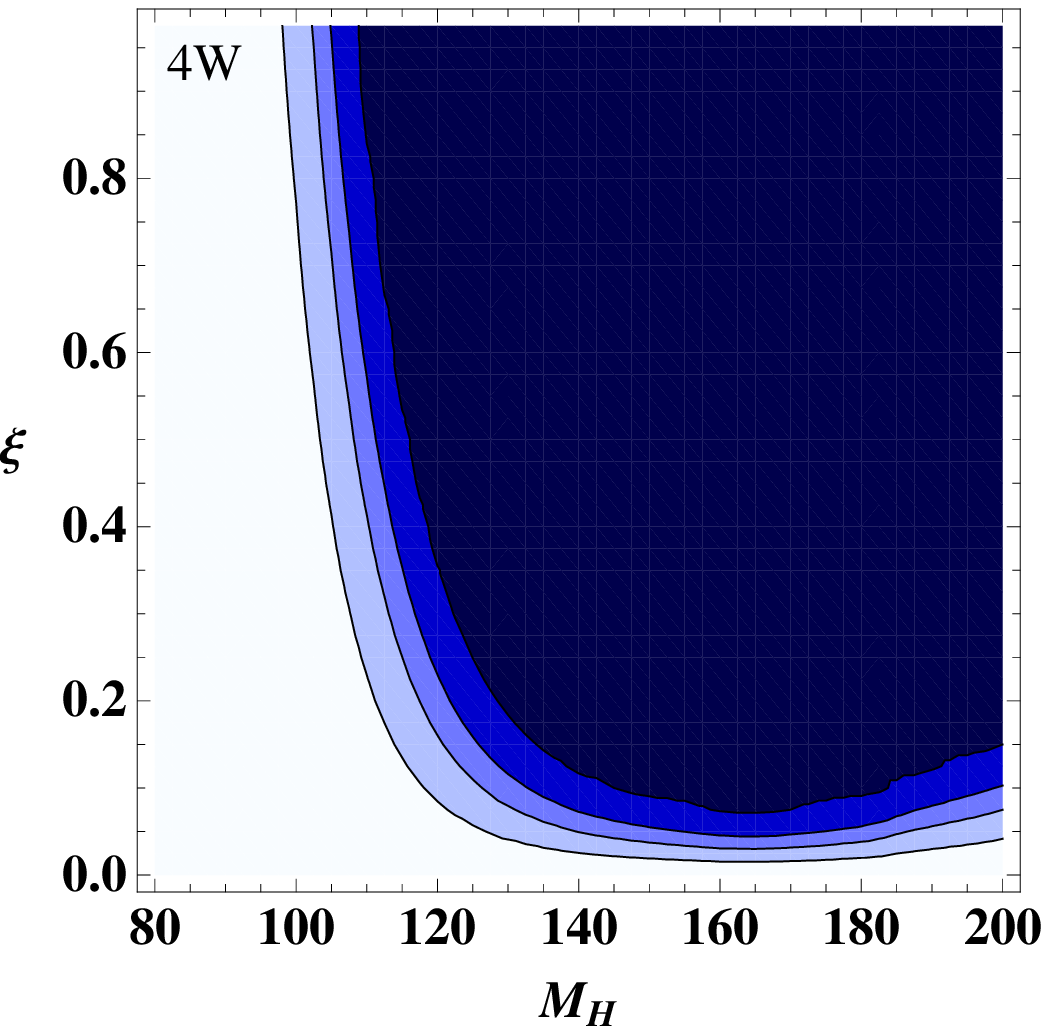,height=3.8cm}
\caption{Areas in the $\xi-M_H$ plane, where in the framework of
  MCHM4 the gluon fusion production of a Higgs pair with subsequent
  decay deviates from the SM. From dark blue (dark gray) to fair blue
  (fair gray) the regions corresponds to 5, 3, 2 and 1$\sigma$. The final states
  are from left to right $b\bar{b}\mu^+\mu^-$, $b\bar{b}\gamma\gamma$,
  $b\bar{b}\tau^+\tau^-$, $4W$, and $\int {\cal L}=300$ fb$^{-1}$.}
\label{fig:dev4}
\end{center}
\vspace*{-0.4cm}
\end{figure}

\begin{figure}[ht]
\vspace*{0.4cm}
\begin{center}
\epsfig{figure=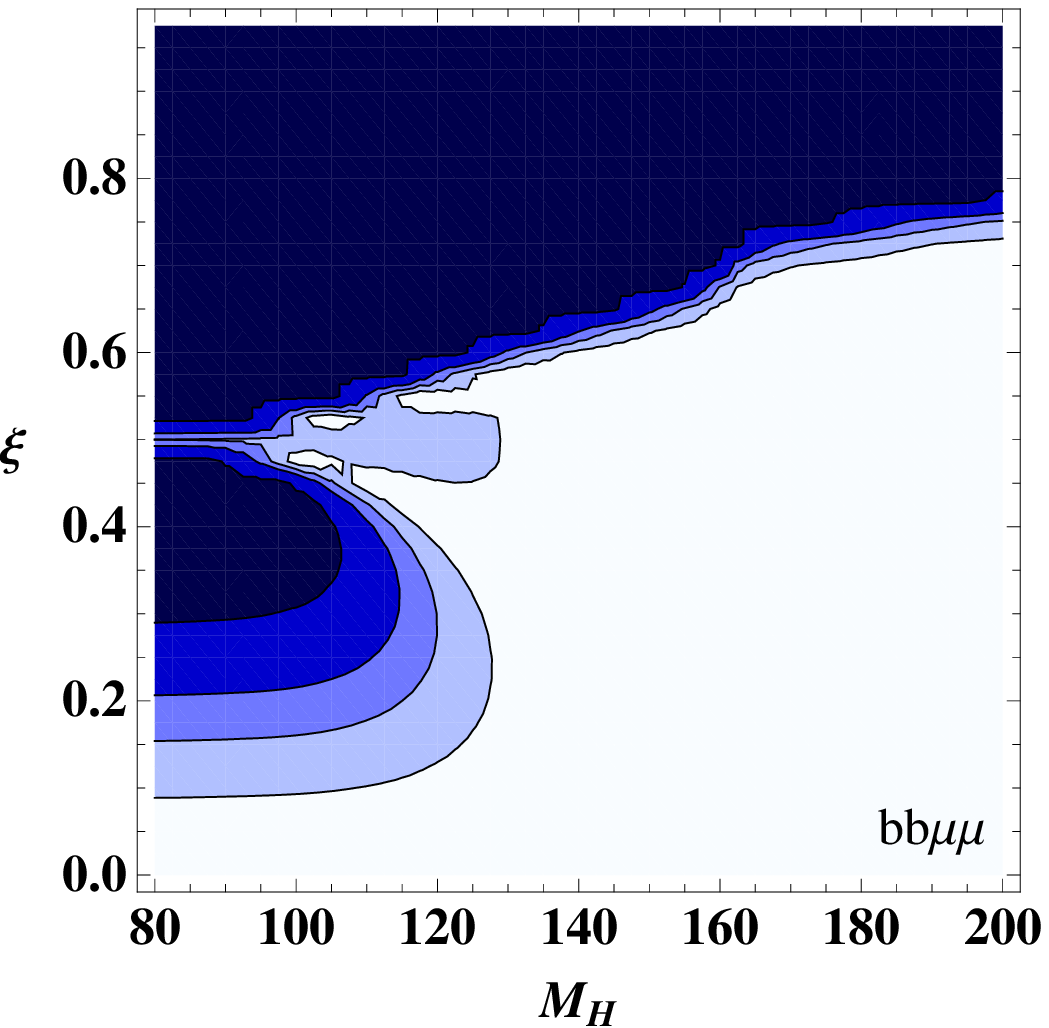,height=3.8cm}
\epsfig{figure=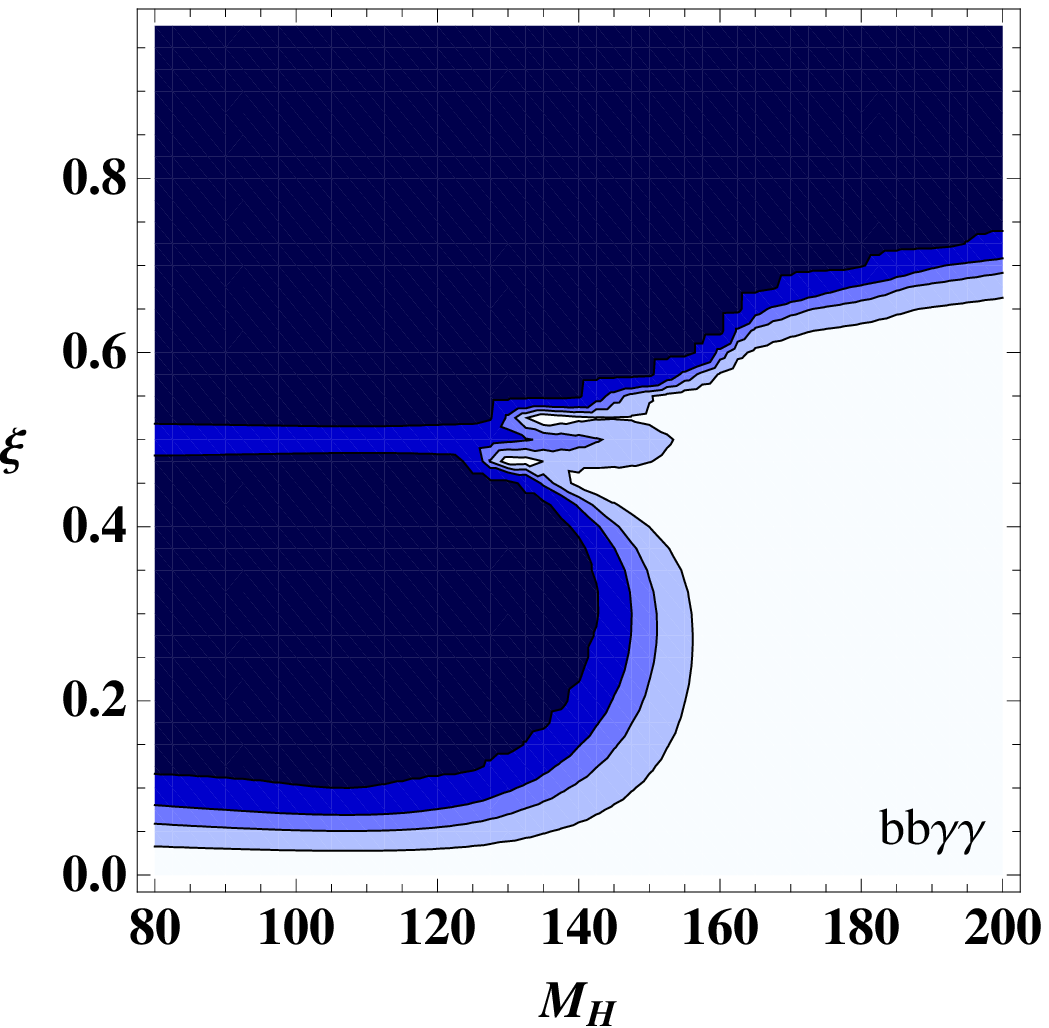,height=3.8cm}
\epsfig{figure=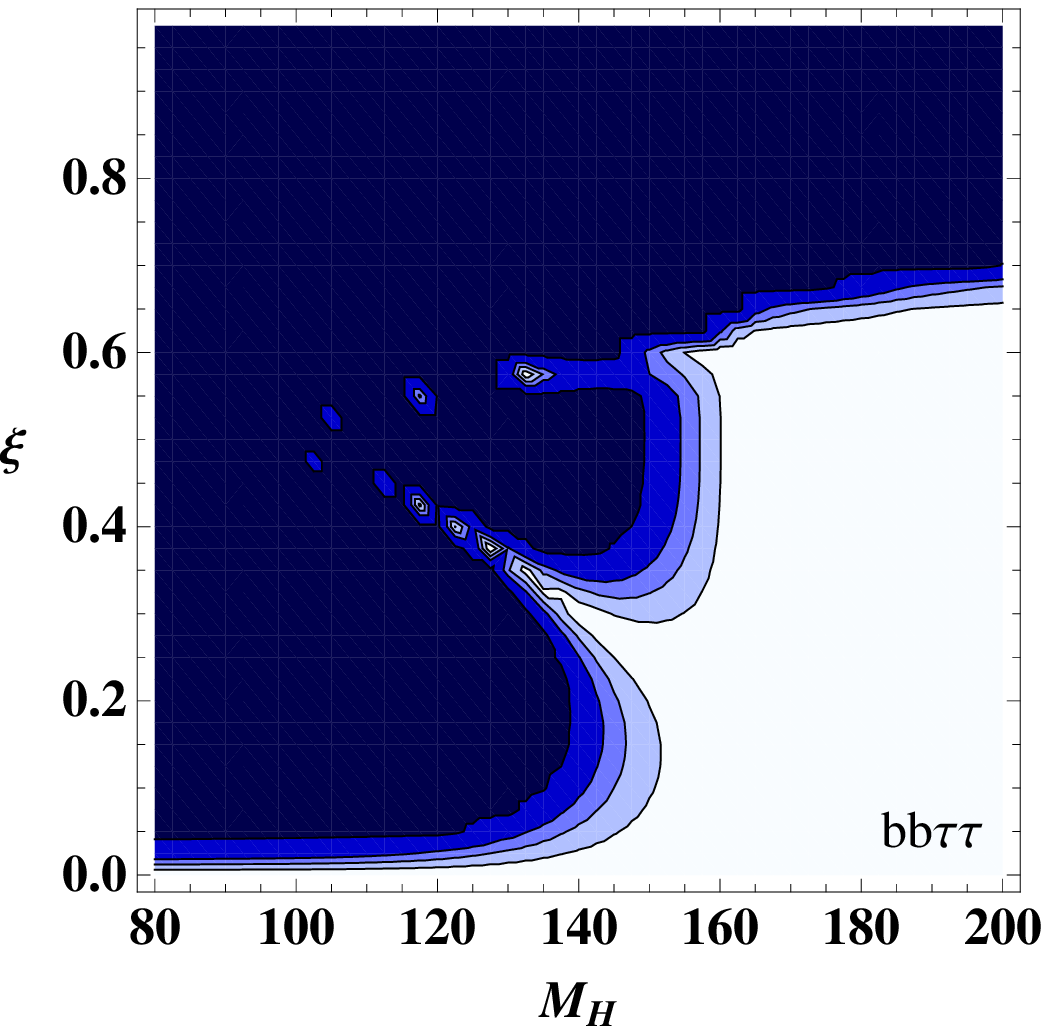,height=3.8cm}
\epsfig{figure=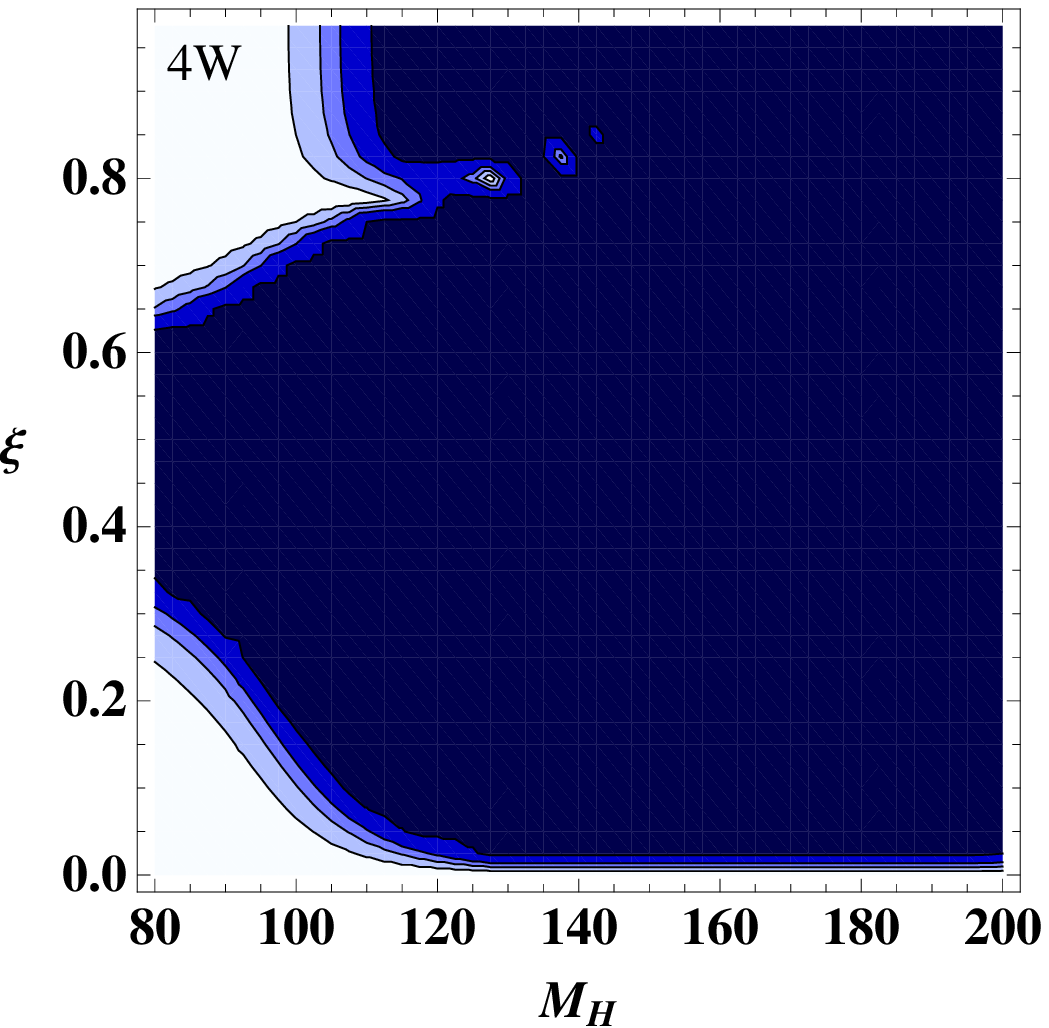,height=3.8cm}
\caption{Areas in the $\xi-M_H$ plane, where in the framework of
  MCHM5 the gluon fusion production of a Higgs pair with subsequent
  decay deviates from the SM. From dark blue (dark gray) to fair blue
  (fair gray) the regions corresponds to 5, 3, 2 and 1$\sigma$. The final states
  are from left to right $b\bar{b}\mu^+\mu^-$, $b\bar{b}\gamma\gamma$,
  $b\bar{b}\tau^+\tau^-$, $4W$, and $\int {\cal L}=300$ fb$^{-1}$.}
\label{fig:dev5}
\end{center}
\vspace*{-0.4cm}
\end{figure}

In order to get a compact answer we have
constructed sensitivity areas in the $\xi-M_H$ plane. 
For each Higgs mass we have determined the value of
$\xi$ where the Higgs pair production cross section with subsequent
decay in the different final states deviates by more than 1, 2, 3 or $5
\sigma$ from the corresponding SM process. Denoting by
$S^{\mbox{\scriptsize MCHM}}$ the number of signal events for the
process calculated in the composite Higgs model and by
$S^{\mbox{\scriptsize SM}}$ the corresponding number in the SM, there
is sensitivity if
\beq
S^{\mbox{\scriptsize SM}} + \beta \, \sqrt{S^{\mbox{\scriptsize
      SM}}} > S^{\mbox{\scriptsize MCHM}} \quad \mbox{or} \quad 
S^{\mbox{\scriptsize SM}} - \beta \, \sqrt{S^{\mbox{\scriptsize
      SM}}} < S^{\mbox{\scriptsize MCHM}} \qquad (\beta=1,2,3,5).
\eeq 
We assumed an integrated luminosity of $300$ fb$^{-1}$.  The results are
presented in Fig.~\ref{fig:dev4} and in Fig.~\ref{fig:dev5} for MCHM4
and MCHM5, respectively. The sensitivity areas are shown for the final states
$b\bar{b}\gamma\gamma$, $b\bar{b}\tau\tau$ and $W^+W^-W^+W^-$. For
comparison we also show $b\bar{b}\mu^+\mu^-$. The sensitivity areas
are a result of an interplay of the production process and the Higgs
branching ratios. \s

In MCHM4, {\it c.f. }Fig.~\ref{fig:dev4}, the deviation from the SM is
exclusively dictated by
the behaviour of the production process, as the MCHM4 branching ratios
are the same as in the SM. In the $b\bar{b}\gamma\gamma$ final state,
MCHM4 can be distinguished from the SM at 3 $\sigma$ 
starting from $\xi \approx 0.6$
and for $M_H \lsim 140$ GeV. Here on the one hand the gluon fusion cross
section and on the other hand the branching ratios are large enough to
yield the necessary number of signal events. The same holds for the
$b\bar{b}\tau\tau$ final state where due to the larger branching
ratio in a $\tau$ pair the $5\sigma$ sensitivity region extends to values of
$\xi$ as low as $\sim 0.2$ for $M_H \lsim 120$~GeV. In the $4W$ final
state the complementary Higgs mass region can be covered, {\it i.e.}
at 5$\sigma$ $M_H \gsim 140$~GeV for $\xi \gsim 0.1$ ($M_H \gsim 110$
GeV for large $\xi$ values).\footnote{The sensitivity areas for the
  $4W$ final state will change, once the $W$ boson decays are included,
  and depend on the final  states into which the $W$ bosons decay.}
The $b\bar{b}\mu\mu$ final state on the other hand is 
  hopeless due to its small branching ratios. The no-sensitivity
  regions are a result of too few events. Altogether, in MCHM4 in
  the whole mass range $\xi$ values above about 0.1 can be tested. As
  expected, for lower $\xi$ values no sensitivity to deviations from
  the SM can be reached.

In MCHM5, the sensitivity reach in the $\xi-M_H$ parameter space is
given by the behaviour of both the Higgs pair production process and
the branching ratios. This is why in the $b\bar{b}\mu\mu$,
$b\bar{b}\gamma\gamma$, $b\bar{b}\tau\tau$ final states the $5\sigma$
sensitivity areas extend to $M_H \sim 200$ GeV for $\xi$ values above about
0.7 in contrast to MCHM4. This is a result of the large $b\bar{b}$
branching ratio which is important up to high mass values for large
$\xi$, {\it c.f.} Figs.~\ref{fig:branch}. The vanishing
branching ratios into the fermionic final states, however, 
result in smaller or even vanishing sensitivity areas
around $\xi=0.5$.\footnote{Note that there is non-vanishing
  sensitivity also for $\xi=0.5$ because we test here sensitivity to
  deviations from the SM and not sensitivity to the Higgs self-coupling.} 
The low-sensitivity regions in the
$b\bar{b}\tau\tau$ final state are due to the same reason: The
fermionic branching ratios vanish at $\xi=0.5$. Furthermore, the
MCHM5 cross section is larger than the SM one for small and large
values of $\xi$ due to the larger production cross section. Therefore
there must be regions below and above $\xi=0.5$ where both the
composite Higgs and the SM rates must be the same so that the
sensitivity vanishes. With a finer grid in $\xi-M_H$ the disconnected
regions of low sensitivity tend to form a connected line extending down to
$M_H=80$ GeV. This behaviour can be clearly seen in the $b\bar{b}\mu\mu$
and $b\bar{b}\gamma\gamma$ final states. The $4W$ final state is
sensitive to SM deviations for all $\xi$ 
values and $M_H \gsim 120$ GeV. For small Higgs masses it also
covers the difficult region around $\xi=0.5$ where the branching ratio
into $W$ bosons is largely enhanced due to the vanishing $b\bar{b}$
branching ratio. The dots of low sensitivity are the result
of an interplay of falling $WW$ branching ratio with rising $\xi$,
see Figs.~\ref{fig:branch}, and a gluon fusion production process which
is not yet large enough to make up for it. In summary, in MCHM5 the
whole $\xi$ mass range for $\xi \gsim 0.05$ can be tested. Compared
to MCHM4, the sensitivity areas are larger for the individual final
states. \s
%\newpage

\noindent
\underline{2.) Is there sensitivity to $\lambda_{HHH}$ in MCHM4 and
  MCHM5?}

\begin{figure}[ht]
\vspace*{0.4cm}
\begin{center}
\epsfig{figure=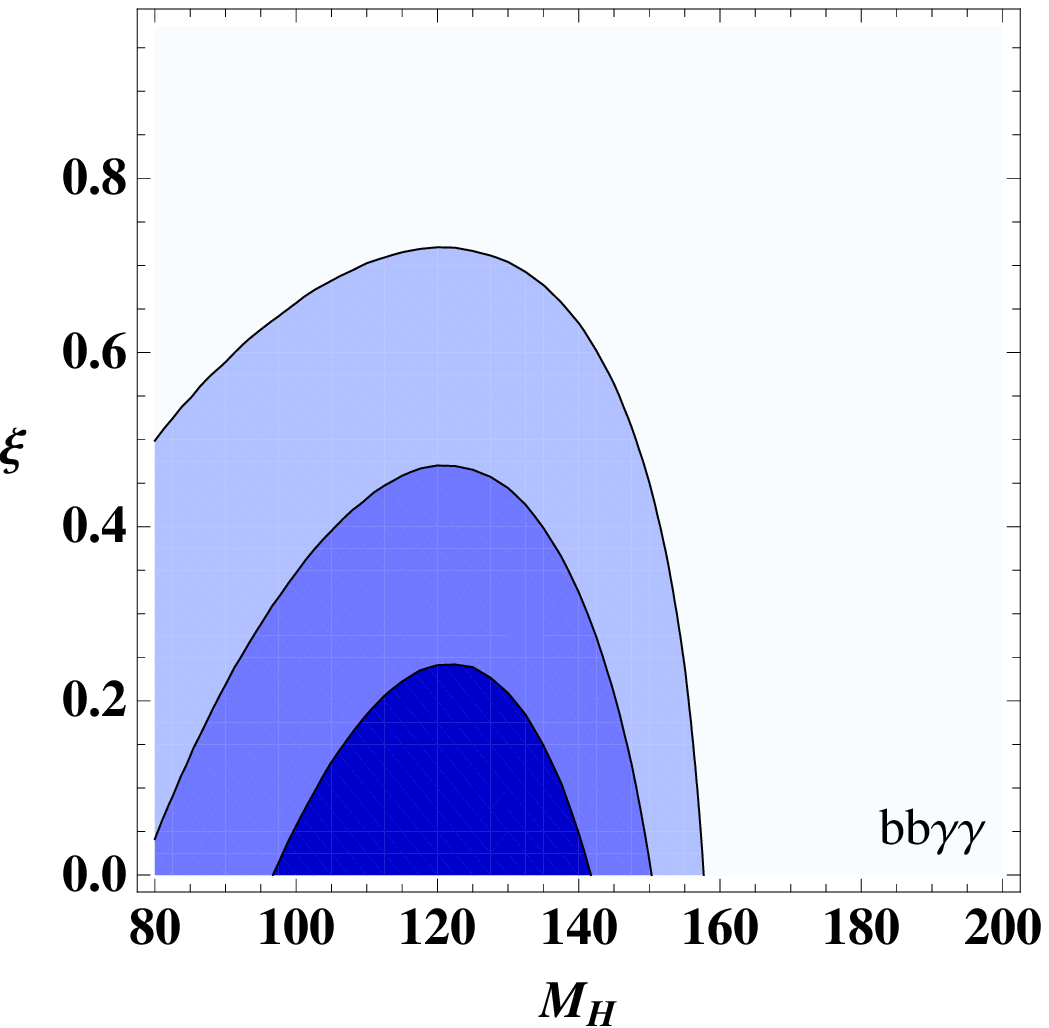,height=5cm}
\epsfig{figure=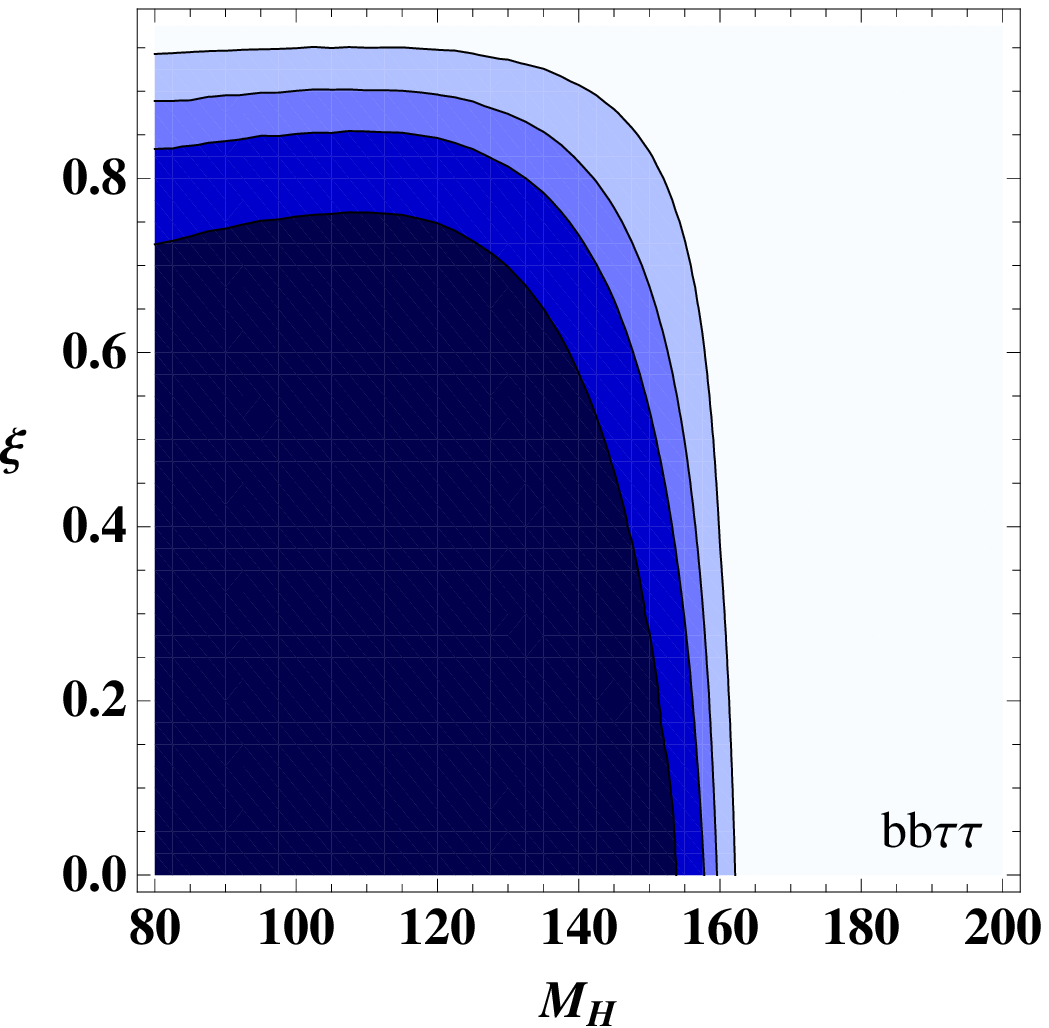,height=5cm}
\epsfig{figure=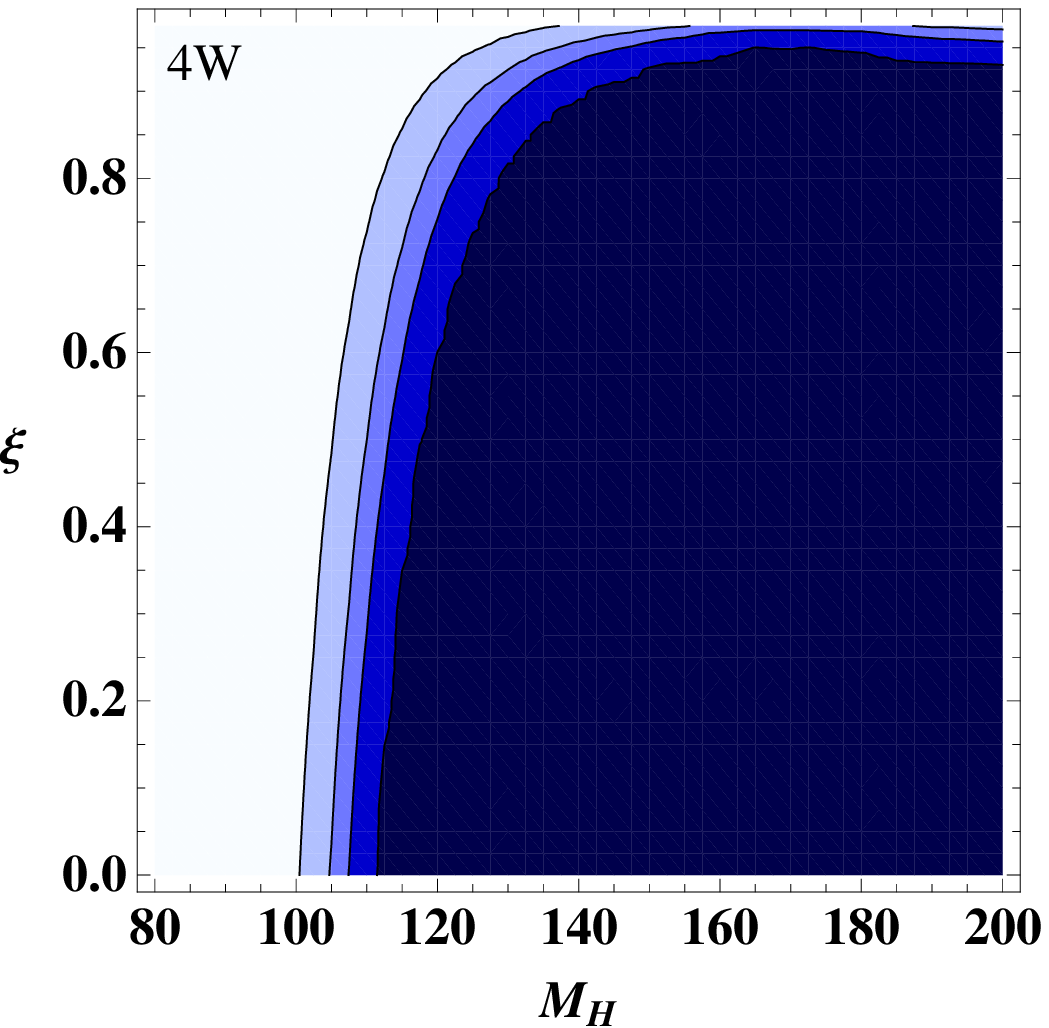,height=5cm}
\caption{Areas in $\xi-M_H$ with sensitivity to non-vanishing
  $\lambda_{HHH}$ in gluon fusion Higgs pair production with
  subsequent decay for MCHM4. From dark blue
  (dark gray) to fair blue (fair gray) the regions correspond to 5, 3,
 2 and 1 $\sigma$. The final states are from left to right
  $b\bar{b}\gamma\gamma$, $b\bar{b}\tau\tau$, $4W$, and $ \int {\cal
    L}=300$ fb$^{-1}$. }
\label{fig:sens4}
\end{center}
\vspace*{-0.4cm}
\end{figure}

\begin{figure}[ht]
\vspace*{0.4cm}
\begin{center}
\epsfig{figure=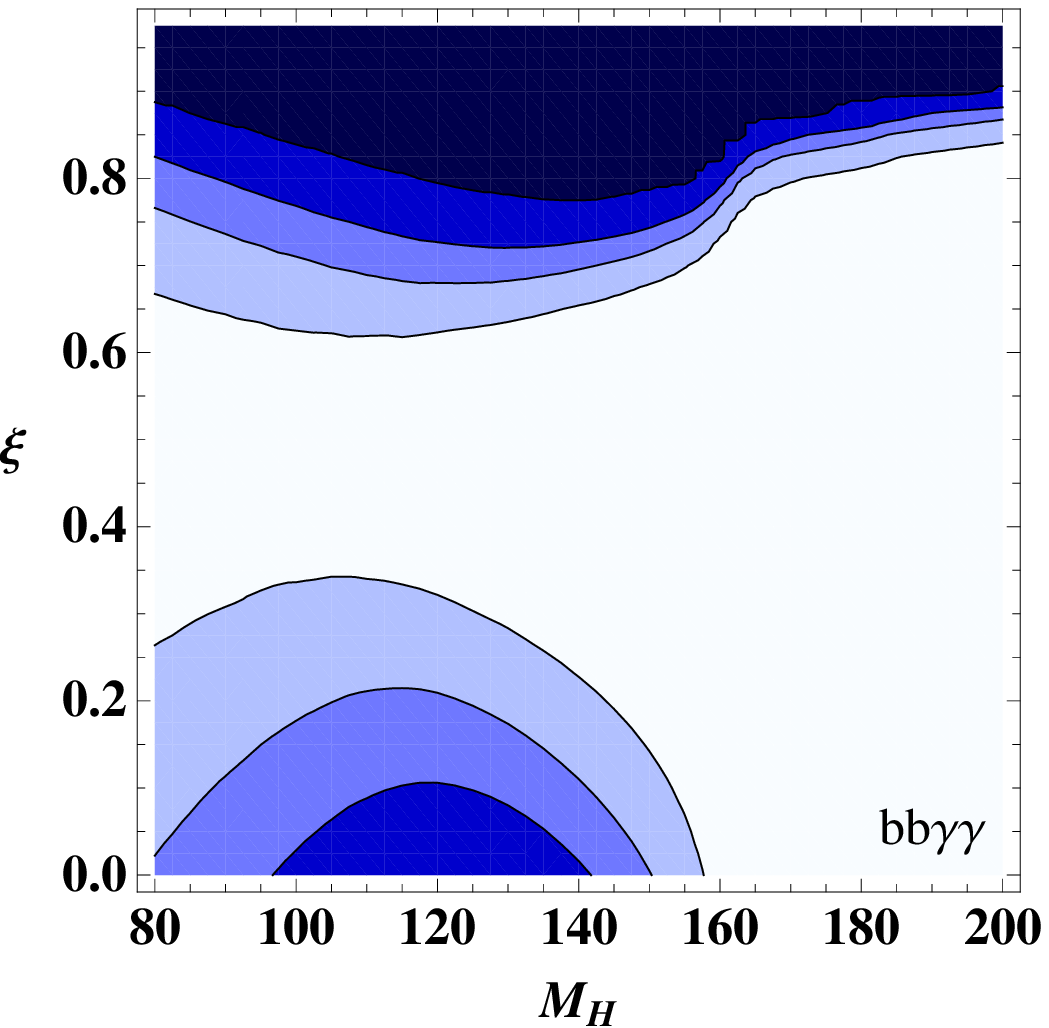,height=5cm}
\epsfig{figure=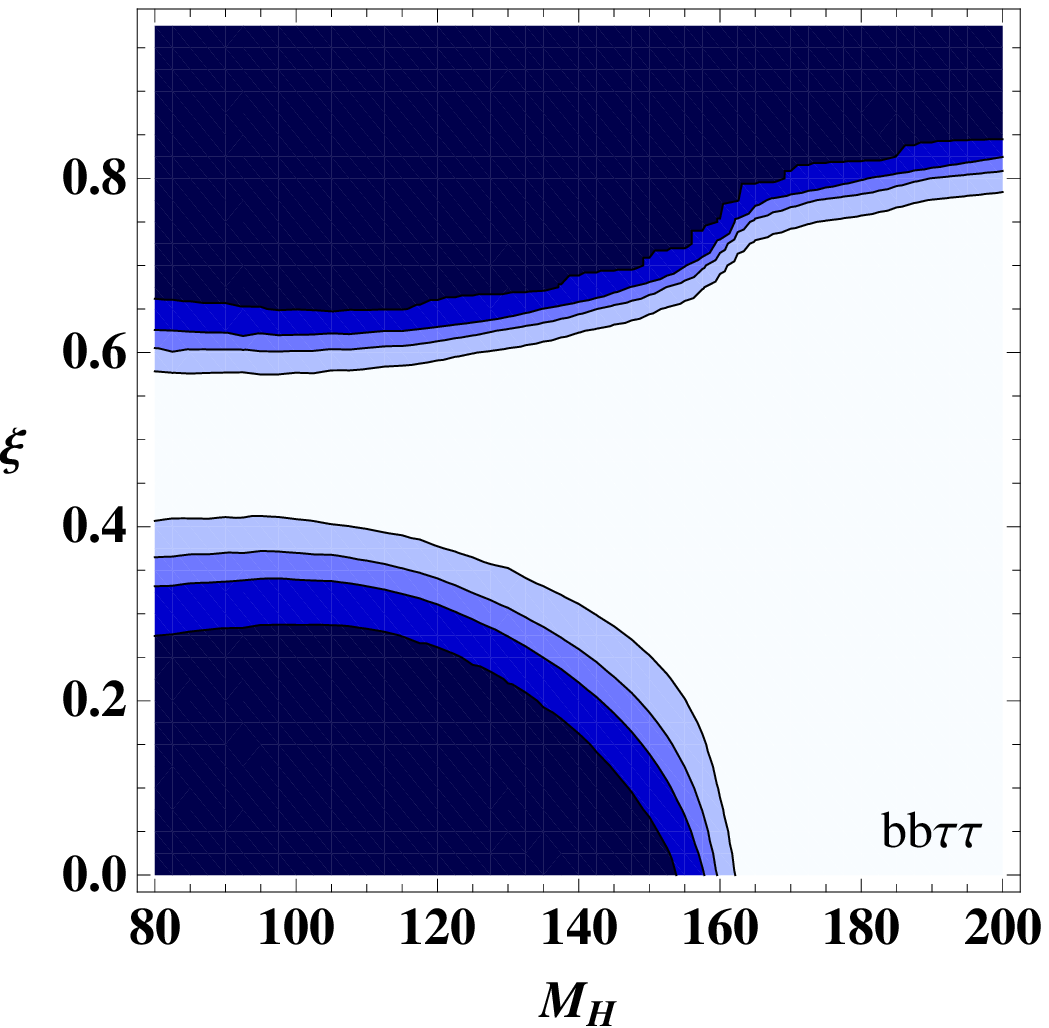,height=5cm}
\epsfig{figure=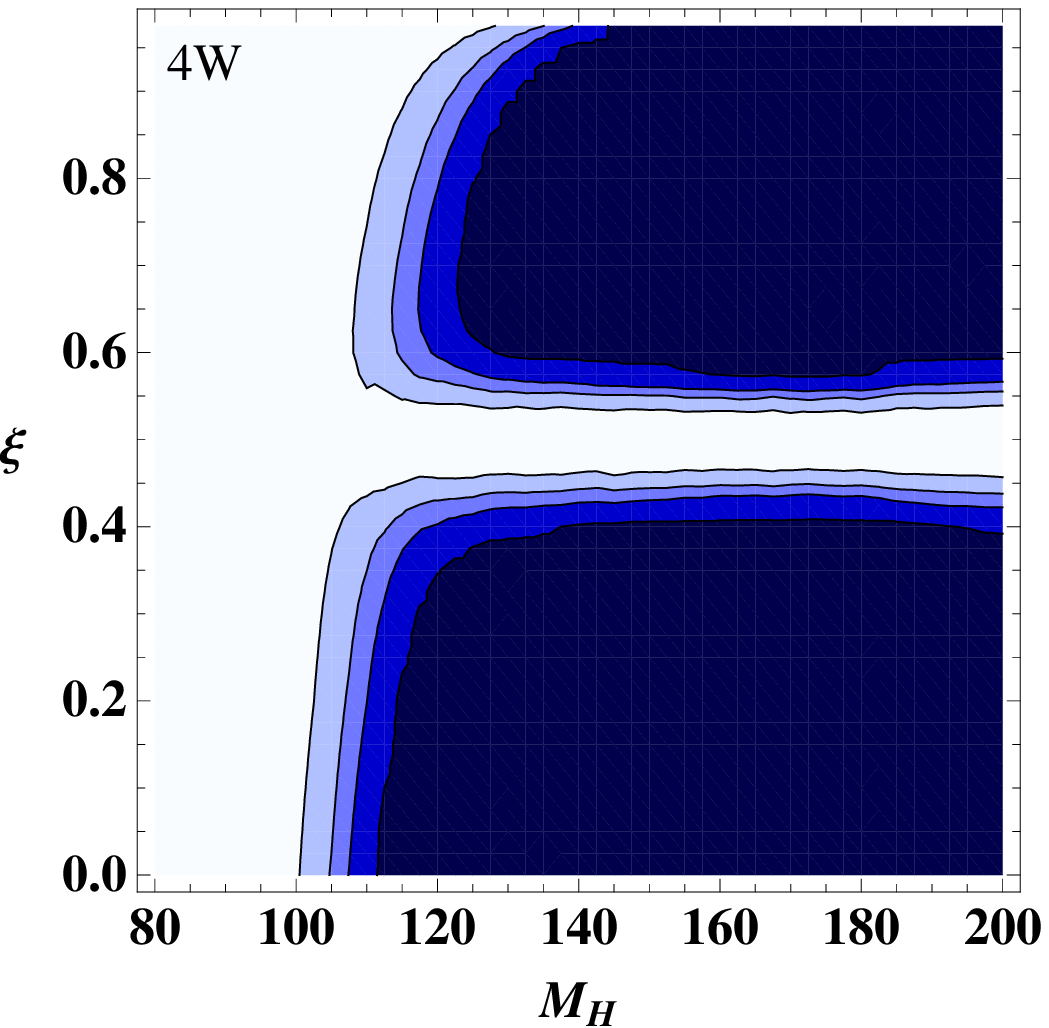,height=5cm}
\caption{Areas in $\xi-M_H$ with sensitivity to non-vanishing
  $\lambda_{HHH}$ in gluon fusion Higgs pair production with
  subsequent decay for MCHM5. From dark blue
  (dark gray) to fair blue (fair gray) the regions correspond to 5, 3,
 2 and 1 $\sigma$. The final states are from left to right
  $b\bar{b}\gamma\gamma$, $b\bar{b}\tau\tau$, $4W$, and $ \int {\cal
    L}=300$ fb$^{-1}$. }
\label{fig:sens5}
\end{center}
\vspace*{-0.4cm}
\end{figure}

Let us now suppose that nature has chosen to realize EWSB in the
framework of a composite Higgs model. Furthermore, we assume that the
Higgs has been discovered and its 
couplings to gauge bosons and fermions are known.\footnote{
Ref.~\cite{Bock:2010nz} analyzed the LHC reach in testing deviations from
the SM Higgs couplings in the context of composite Higgs
models. It was shown that $\xi$ can be extracted with an accuracy of
${\cal O}(20\%)$.} We want to investigate if a non-zero
$\lambda_{HHH}$ coupling can be established from the gluon fusion
Higgs pair production process so that the Higgs mechanism can be 
corroborated experimentally. To this goal we derived sensitivity
areas in the $\xi-M_H$ parameter space where the cross section with
vanishing trilinear Higgs coupling deviates by more than 1, 2, 3 and
$5\sigma$ at $\int \, {\cal L}= 300$ fb$^{-1}$ from the composite
Higgs process with non-zero Higgs self-interaction strength, {\it
  i.e.} as given in Eqs.~(\ref{eq:self4},\ref{eq:self5}). They are presented in
Fig.~\ref{fig:sens4} and Fig.~\ref{fig:sens5} for MCHM4 and MCHM5. 
The presented final states are 
$b\bar{b}\gamma\gamma$, $b\bar{b}\tau\tau$ and $4W$.\footnote{We do
  not show the sensitivity area in the $b\bar{b}\mu\mu$ final state, since
  it is too small.}  

In MCHM4, the contours of the sensitivity areas follow the behaviour
of the branching ratios along the $M_H$ direction, {\it
  c.f.}~Fig.~\ref{fig:smbranch}. With decreasing importance of the
respective branching ratio the sensitivity to non-zero
$\lambda_{HHH}$ drops. In the $\xi$ direction the sensitivity is
dominated by the production process. With rising $\xi$ the sensitivity
in gluon fusion diminishes, as the diagram with
direct coupling of a Higgs to a fermion pair, which is linear in
$\xi$, becomes more and more important. This has been discussed in
detail in Section \ref{sec:hhcxn}. Note that the sensitivity area in the
$b\bar{b}\gamma\gamma$ final state where the composite Higgs model can
be distinguished from the SM is complementary to the one where a
non-zero trilinear Higgs coupling can be established. In the
$b\bar{b}\tau\tau$ and $4W$ final states, however, there is
considerable overlap of these regions. \s

In MCHM5,  in the $b\bar{b} \gamma\gamma$ and
$b\bar{b}\tau\tau$ final states there is sensitivity to
$\lambda_{HHH}$ also for large $\xi$ values up to $M_H=200$ GeV. This
is because  the fermionic branching ratios 
stay important up to large Higgs masses for $\xi> 0.5$.
For $\xi=0.5$, however, $\lambda_{HHH}$ vanishes so that around this
value there is no sensitivity at all. In the same final states for low
$\xi$ values the sensitivity areas are smaller than in MCHM4, since
the trilinear and Yukawa couplings are more strongly suppressed here
compared to MCHM4. \s

\begin{figure}[ht]
\vspace*{0.4cm}
\begin{center}
\epsfig{figure=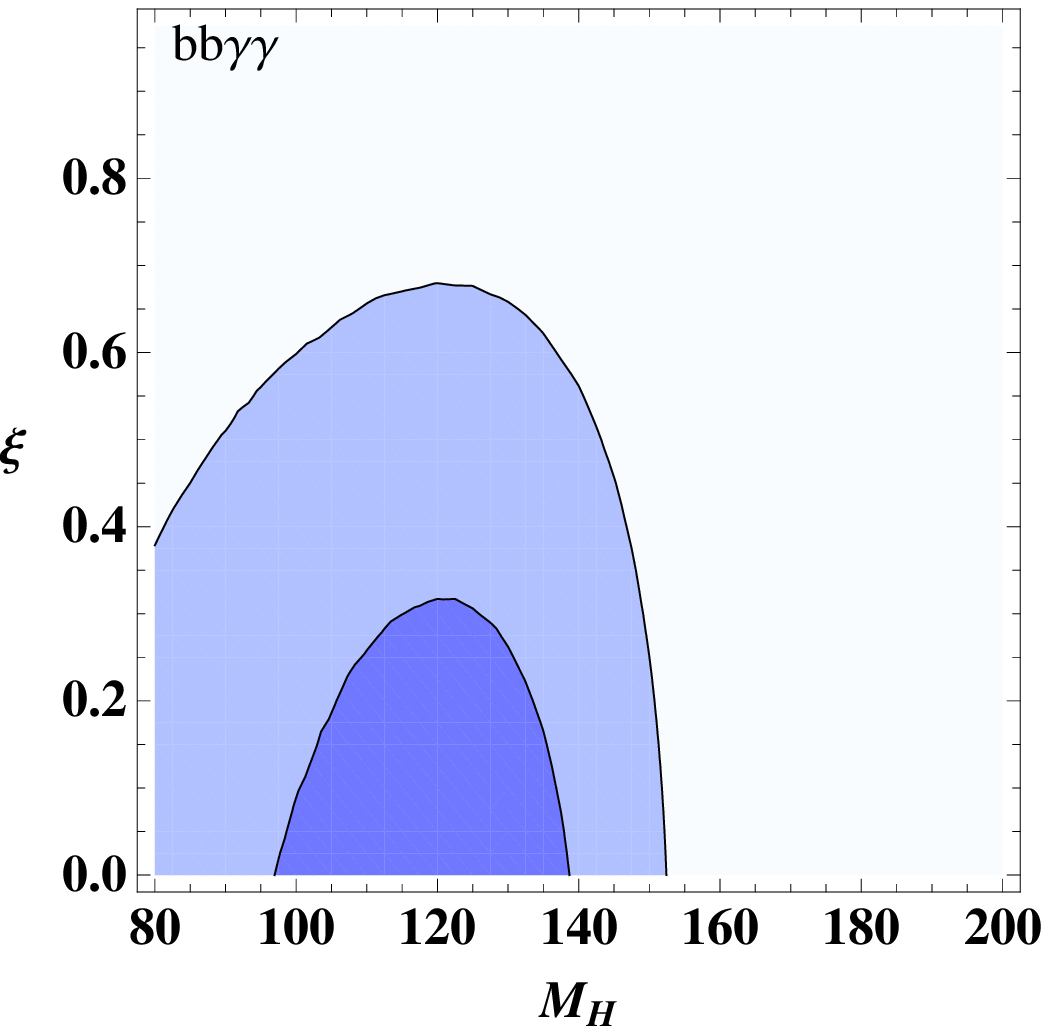,height=5cm}
\epsfig{figure=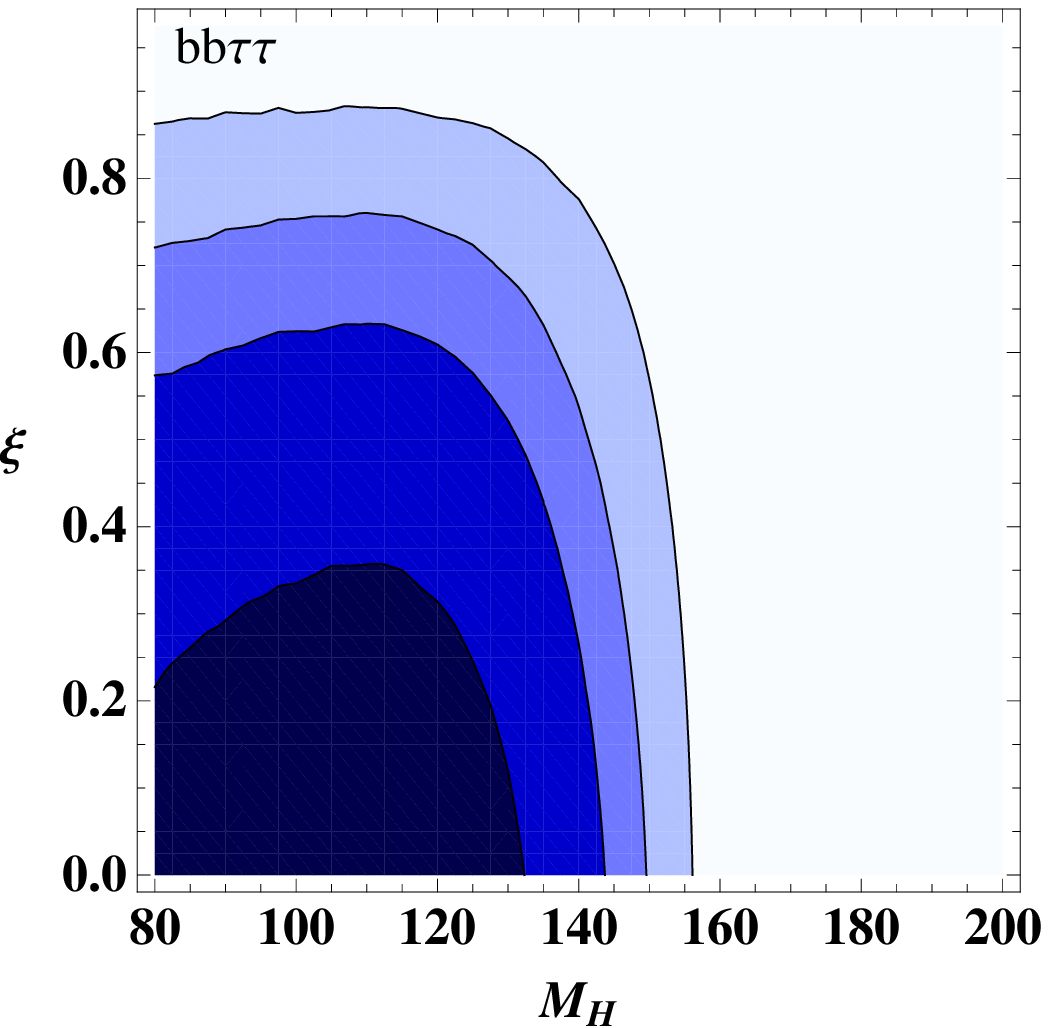,height=5cm}
\epsfig{figure=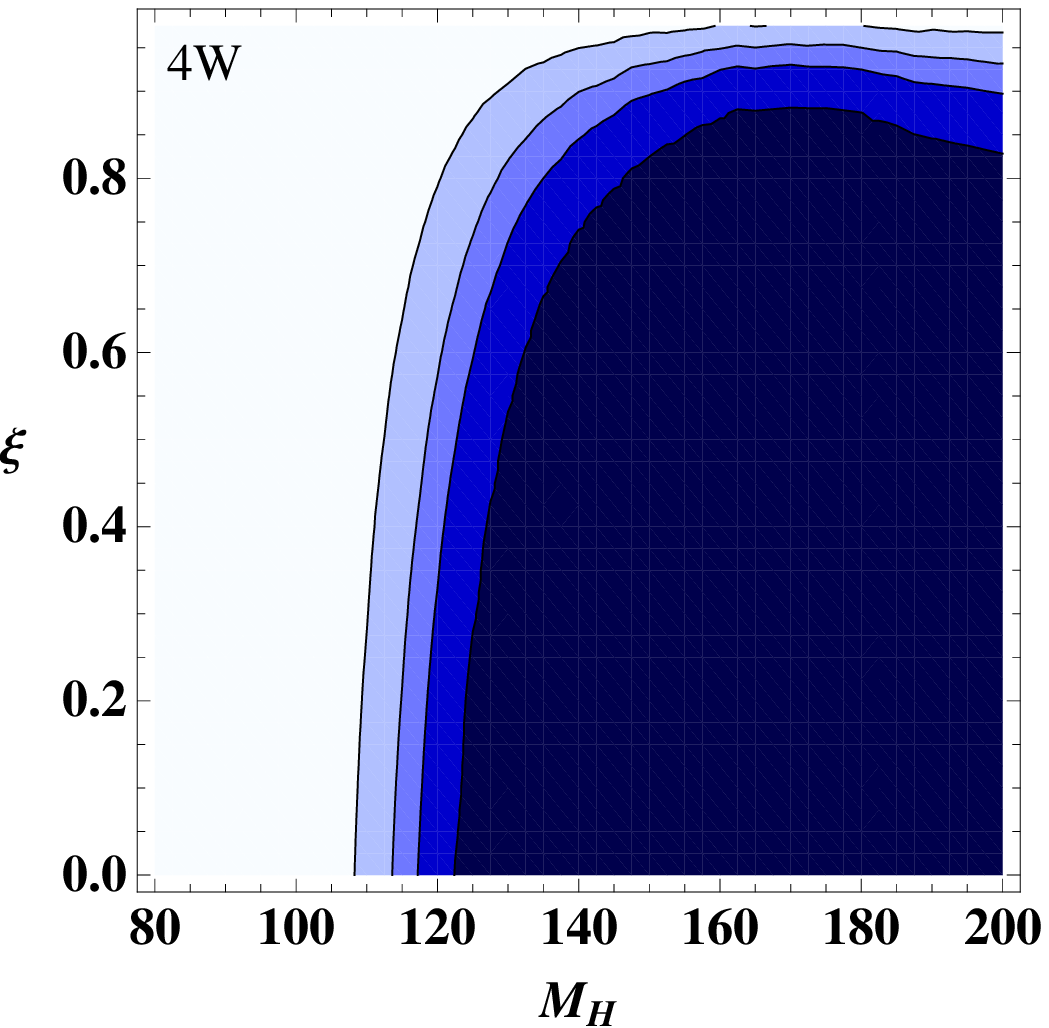,height=5cm}
\caption{Areas in $\xi-M_H$ with sensitivity to a $\lambda_{HHH}$
  variation of 30\% in gluon fusion Higgs pair production with
  subsequent decay for MCHM4. From dark blue
  (dark gray) to fair blue (fair gray) the regions correspond to 5, 3,
 2 and 1 $\sigma$. The final states are from left to right
  $b\bar{b}\gamma\gamma$ at $ \int {\cal
    L}=3000$ fb$^{-1}$ and $b\bar{b}\tau\tau$, $4W$ at $ \int {\cal
    L}=600$ fb$^{-1}$. }
\label{fig:mchm4sens30}
\end{center}
\vspace*{-0cm}
\end{figure}

\begin{figure}[ht]
\vspace*{0.4cm}
\begin{center}
\epsfig{figure=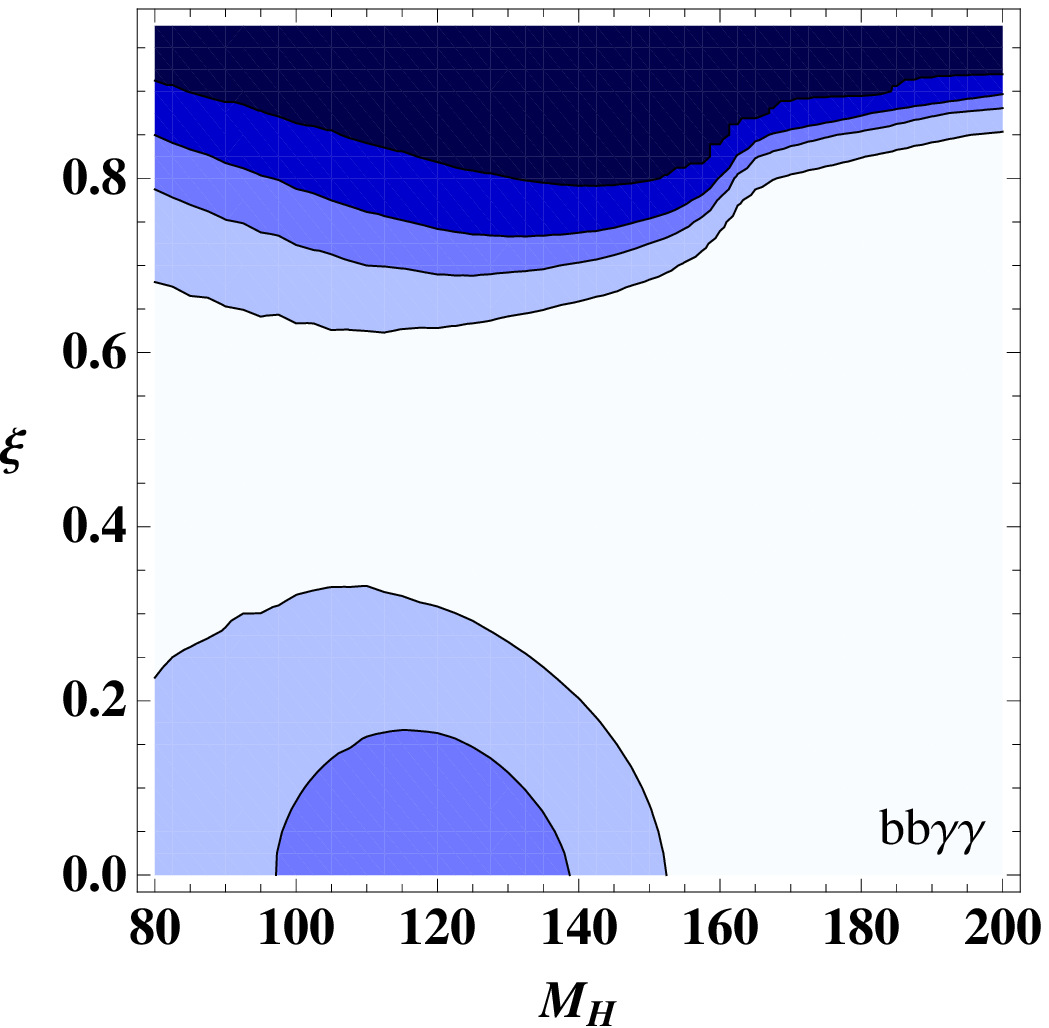,height=5cm}
\epsfig{figure=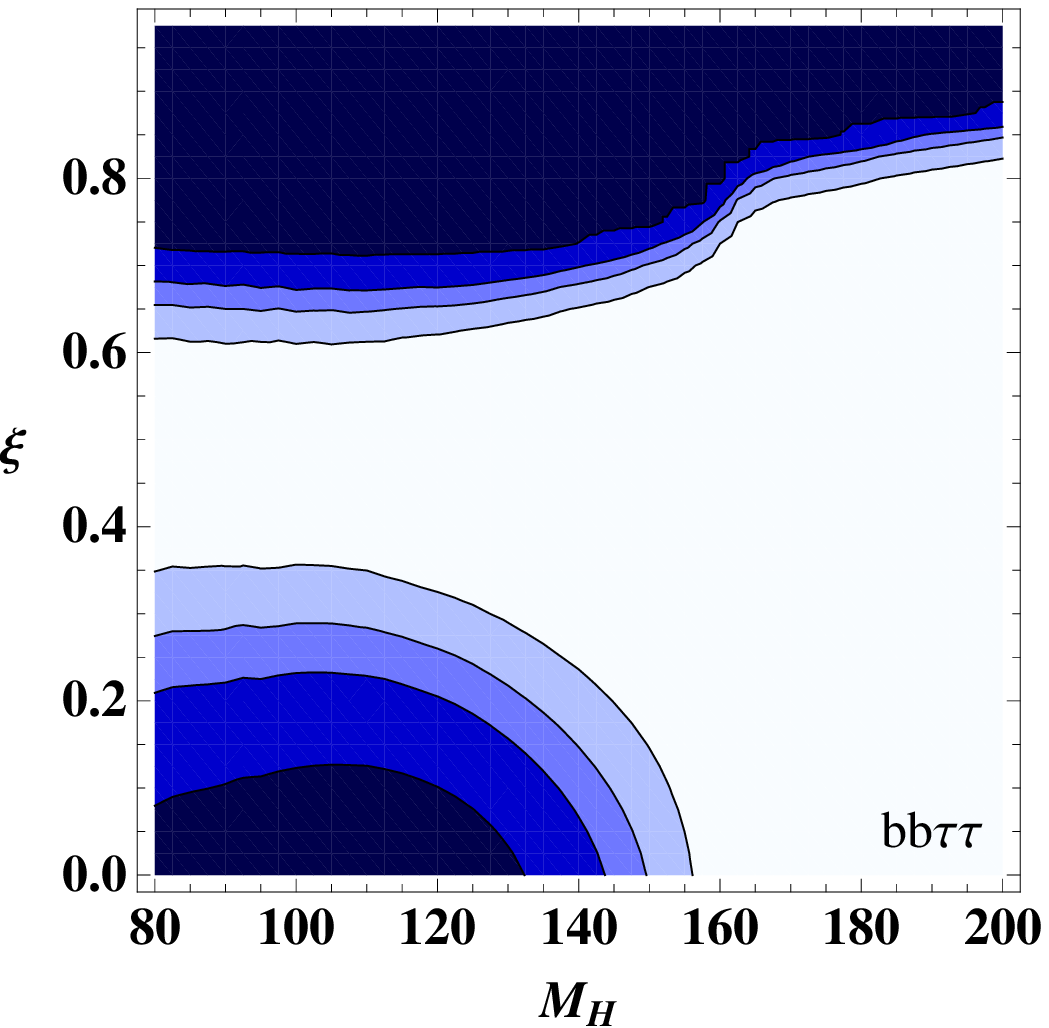,height=5cm}
\epsfig{figure=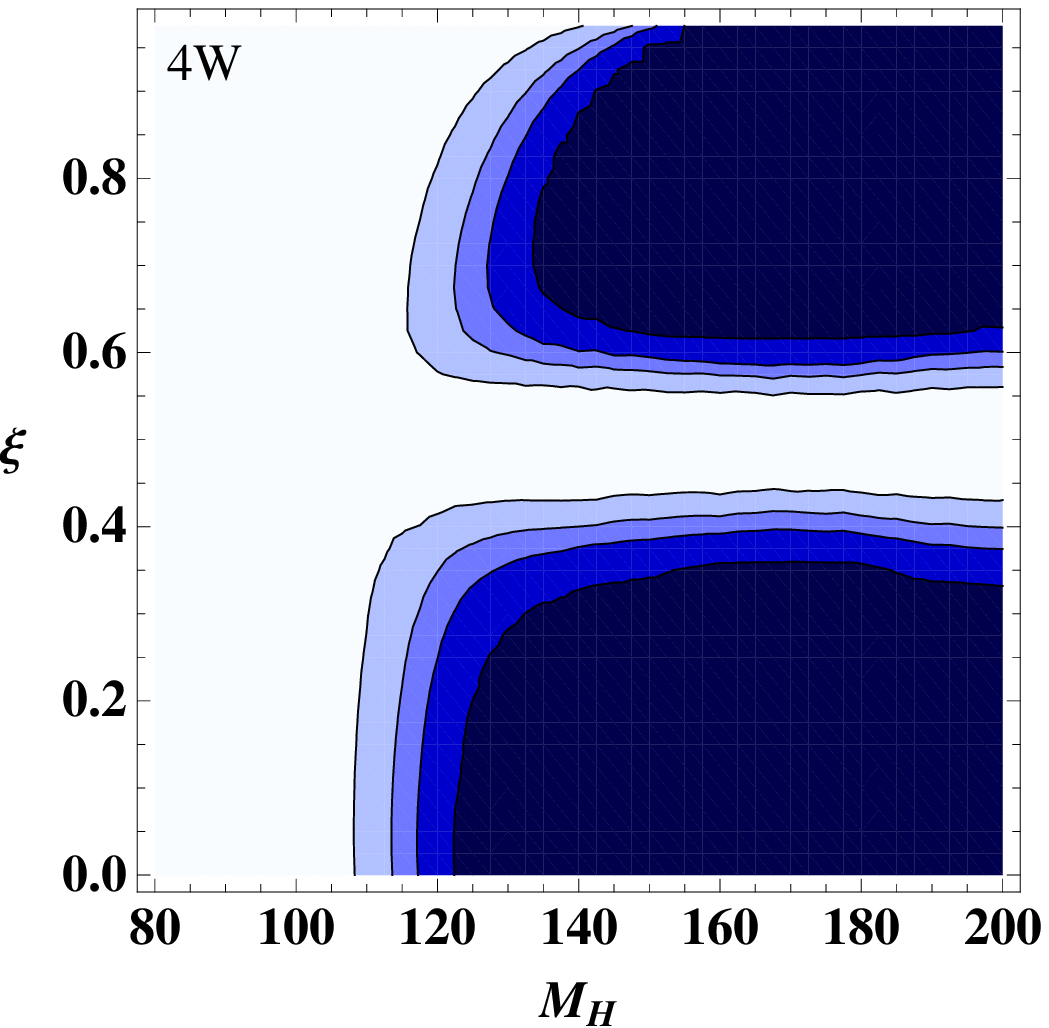,height=5cm}
\caption{Areas in $\xi-M_H$ with sensitivity to a $\lambda_{HHH}$
  variation of 30\% in gluon fusion Higgs pair production with
  subsequent decay for MCHM5. From dark blue
  (dark gray) to fair blue (fair gray) the regions correspond to 5, 3,
 2 and 1 $\sigma$. The final states are from left to right
  $b\bar{b}\gamma\gamma$ at $ \int {\cal
    L}=3000$ fb$^{-1}$ and $b\bar{b}\tau\tau$, $4W$ at $ \int {\cal
    L}=600$ fb$^{-1}$. }
\label{fig:mchm5sens30}
\end{center}
\vspace*{-0.4cm}
\end{figure}

Finally, we show in Figs.~\ref{fig:mchm4sens30} and
\ref{fig:mchm5sens30} the sensitivity to a variation of
$\lambda_{HHH}$ by $+30$\%, {\it i.e.}
\beq
\begin{array}{lcll}
\lambda'_{HHH} &=& 1.3 \, \sqrt{1-\xi} 
\lambda_{HHH}^{\mbox{\scriptsize SM}} &
\qquad \mbox{for MCHM4} \\[0.2cm]
\lambda'_{HHH} &=& 1.3 \, \frac{1-2\xi}{\sqrt{1-\xi}}
\lambda_{HHH}^{\mbox{\scriptsize SM}} &
\qquad \mbox{for MCHM5} \;.
\end{array}
\eeq
 We have chosen an integrated luminosity
of 600 fb$^{-1}$. This corresponds to 3 years of running at the LHC
design luminosity with two detectors. As expected the sensitivity
regions shrink considerably. Nevertheless, in the $4W$ final state
above $\sim 120$ GeV a large part of the $\xi$ range can be
covered. For low Higgs masses only the difficult $b\bar{b}\tau\tau$
final state allows for a 30\% determination of $\lambda_{HHH}$ at low
$\xi$ values. In the $b\bar{b}\gamma\gamma$ final state the high
integrated luminosity of an SLHC with 3000 fb$^{-1}$ would be needed.\s

\noindent
\underline{Feasibility:}
Of course the sensitivity areas shown in Figs. 8-11 represent the ideal case,
where we assume that the underlying composite Higgs model has already been
pinned down with high accuracy. Furthermore, we assumed the double Higgs
production cross sections to be large enough to be measurable. 
The question arises as to what extent the sensitivity areas will shrink in a 
full analysis, taking into account the background reactions in a systematic
way as well as detector properties.\footnote{For parton level analyses
  in the SM see Refs.\cite{Lafaye:2000ec,Baur:2002rb,Baur:2003gpa}.}
Although this is beyond the scope of this work we can
discuss this question qualitatively. As example let us look at the
$4W$ final state. In Refs.~\cite{Baur:2002rb} such an analysis has
been performed for the SM Higgs pair production in gluon fusion with
subsequent decay of the $W$ boson pairs 
into four jets and two same-sign leptons, respectively. The main
background originates from $W^\pm W^+W^-jj$ production followed by the
$t\bar{t}W^\pm$ background where one top quark decays leptonically, the other
hadronically. It has been found that for 300 fb$^{-1}$ a vanishing
Higgs self-coupling can be excluded at the 95\% CL or better in the
mass range 150 to 200 GeV. Furthermore, at the SLHC with 3000
fb$^{-1}$ $\lambda_{HHH}$ can be determined with an accuracy of
20-30\% for $160 \le M_H \le 180$ GeV. If we look at this in the
framework of the composite Higgs model the backgrounds do not change
as long as they do not involve any Higgs intermediate state. Such a
process would be the subleading electroweak process of Higgs-strahlung off a
$W$ boson with subsequent decay into $W^+W^-$. Since in MCHM4 and MCHM5 the Higgs couplings to gauge bosons
are always suppressed  compared to the SM we are conservative if we
compare to the SM background. A Higgs produced in Higgs-strahlung with
subsequent decay  into $t\bar{t}$ does not present any danger, since
we are well below the top quark pair threshold. As for the signal
process in MCHM4 the $HH$ production cross section in gluon fusion
with subsequent 
decay in 4$W$ bosons exceeds the SM process for $\xi$ values up to
about 0.7. In MCHM5 with an even more enhanced production cross
section this regions extends up to $\xi$ close to 1, where it finally
vanishes due to zero Higgs couplings to gauge bosons. Combined with
our knowledge about the sensitivity areas presented in this section we
can conclude that the prospects of excluding vanishing $\lambda_{HHH}$
or even measuring it with about 30\% accuracy are encouraging. If the
Higgs mass is below about 140 GeV the situation becomes more
difficult. One would have to exploit final states with $b$-quarks and
photons or $\tau$ leptons. The extraction of $\lambda_{HHH}$ is much
more difficult here, as has been shown for the SM case in
\cite{Baur:2003gpa}. In MCHM5 we have the additional complication of
not being sensitive at all to $\lambda_{HHH}$ for $\xi \approx 0.5$.

%%%%%%%%%%%%%%%%%%%%%%%%%%%%%%%%%%%%%%%%%%%%%%%%%%%%%%%

\section{Summary} \label{sec:concl}
We have systematically investigated the Higgs pair production
processes in the minimal 
composite Higgs models MCHM4 and MCHM5, which give access to the
trilinear Higgs self-coupling. A measurement of $\lambda_{HHH}$ allows
to make a first step towards the reconstruction of the Higgs
potential. Furthermore, it provides information on the dynamics beyond
EWSB. Due to the small cross sections and large backgrounds the
measurement is very challenging. For various final states 
we have constructed areas
in the $\xi-M_H$ plane with sensitivity to
deviations from the SM as well as sensitivity to a non-vanishing
trilinear Higgs self-coupling. As for the $4W$ final state the results
are very encouraging and may trigger more sophisticated analyses
taking into account all backgrounds and detector effects, which is
beyond the scope of this paper.

%%%%%%%%%%%%%%%%%%%%%%%%%%%%%%%%%%%%%%%%%%%%%%%%%%%%%%%

\section*{Acknowledgments}

\noindent
We would like to thank Jos\'e Espinosa, Christophe Grojean, Tilman
Plehn, Michael Rauch, Heidi Rzehak and Michael Spira for helpful
discussions. Furthermore, we are grateful to J. Espinosa,
Ch. Grojean and M. Spira for the careful reading of the
manuscript. This research was
supported in part by the Deutsche Forschungsgemeinschaft via the
Sonderforschungsbereich/Transregio SFB/TR-9 Computational Particle
Physics. MMM would like to thank CERN for kind hospitality where part
of this work has been performed.

\end{document}